\documentclass[traditabstract,longauth]{aa} % for the abstract without structuration 
\usepackage{graphicx}
\usepackage{txfonts}
\usepackage{natbib}
\bibpunct{(}{)}{;}{a}{}{,}

\begin{document}

   \title{Ephemeris refinement of 21 hot Jupiter exoplanets with high timing uncertainties}

   \titlerunning{Refinement of hot Jupiter ephemerides}

   \author{M.~Mallonn\inst{1}, C.~von Essen\inst{2}, E.~Herrero\inst{3,4},  X.~Alexoudi\inst{1}, T.~Granzer\inst{1}, M.~Sosa\inst{5,6}, K.\,G.~Strassmeier\inst{1}, G.~Bakos\inst{7,33}, D.~Bayliss\inst{8}, R.~Brahm\inst{9,10,11}, M.~Bretton\inst{12}, F.~Campos\inst{13}, L.~Carone\inst{14}, K.\,D.~Col\'{o}n\inst{15}, H.\,A.~Dale\inst{16},  D.~Dragomir\inst{17,34}, N.~Espinoza\inst{14,35,36}, P.~Evans\inst{18}, F.~Garcia\inst{19}, S.-H.~Gu\inst{20,21,22}, P.~Guerra\inst{23},  Y.~Jongen\inst{24}, A.~Jord\'an\inst{9,11,14}, W.~Kang\inst{25}, E.~Keles\inst{1}, T.~Kim\inst{25}, M.~Lendl\inst{26,14}, D.~Molina\inst{27}, M.~Salisbury\inst{28}, F.~Scaggiante\inst{29}, A.~Shporer\inst{17}, R.~Siverd\inst{30}, E.~Sokov\inst{31,32}, I.~Sokova\inst{32}, A.~W\"{u}nsche\inst{12}}
   \authorrunning{M. Mallonn et al.}

\institute{Leibniz-Institut f\"{u}r Astrophysik Potsdam, An der Sternwarte 16, D-14482 Potsdam, Germany \\ 
  \email{mmallonn@aip.de}
\and 
Stellar Astrophysics Centre, Department of Physics and Astronomy, Aarhus University, Ny Munkegade 120, DK-8000 Aarhus C, Denmark 
\and 
Institut de Ci\'{e}ncies de l'Espai (ICE, CSIC), Campus UAB, Carrer de Can Magrans s\/n, 08193 Cerdanyola del Vall\`{e}s, Spain 
\and
Institut d’Estudis Espacials de Catalunya (IEEC), C/Gran Capit\'{a} 2-4, Edif. Nexus, 08034 Barcelona, Spain
\and
Facultad de Ciencias Astron\'omicas y Geof\'{\i}sicas, Universidad Nacional de La Plata, Paseo del Bosque, B1900FWA, La Plata, Argentina
\and
Instituto de Astrof\'{\i}sica de La Plata (CCT-La Plata, CONICET-UNLP), Paseo del Bosque, B1900FWA, La Plata, Argentina
\and
Department of Astrophysical Sciences, Princeton University, 4 Ivy Lane, Princeton, NJ 08544, US
\and
Dept. of Physics, University of Warwick, Gibbet Hill Road, Coventry CV4 7AL, UK
\and
Center of Astro-Engineering UC, Pontificia Universidad Cat\'olica de Chile, Av. Vicu\~{n}a Mackenna 4860, 7820436 Macul, Santiago, Chile
\and
Instituto de Astrof\'isica, Facultad de F\'isica, Pontificia Universidad Cat\'olica de Chile, Av. Vicu\~{n}a Mackenna 4860, 7820436 Macul, Santiago, Chile
\and
Millennium Institute of Astrophysics, Av. Vicu\~na Mackenna 4860, 782-0436 Macul, Santiago, Chile
\and
Observatoire des Baronnies provencales, Observatoire Astronomique, 05150 Moydans, France
\and
Observatory Puig d'Agulles, Carrer Riera 1, 08759 Vallirana, Spain
\and
Max-Planck-Institut f\"ur Astronomie, K\"onigstuhl 17, 69117 Heidelberg, Germany
\and
NASA Goddard Space Flight Center, Exoplanets and Stellar Astrophysics Laboratory, Greenbelt, MD 20771, US
\and
Emory University Department of Physics, 400 Dowman Drive, Suite N218, Atlanta GA 30322, US
\and
Kavli Institute for Astrophysics and Space Research, Massachusetts Institute of Technology, Cambridge, MA 02139, US
\and
Rarotonga Observatory, PO Box 876, Rarotonga, Cook Islands
\and
La Vara, Valdes Observatory, 33784 Mu\~{n}as de Arriba, Vald\'{e}s, Asturias, Spain
\and
Yunnan Observatories, Chinese Academy of Sciences, Kunming 650011, PR China
\and
Key Laboratory for the Structure and Evolution of Celestial Objects, Chinese Academy of Sciences, Kunming 650216, China
\and
University of Chinese Academy of Sciences, Beijing 100049, China
\and
Observatori Astron\`{o}mic Albany\`{a}, Camí de Bassegoda s/n, 17733 Albany\`{a}, Spain
\and
Observatoire Astronomique de Vaison, 84110 Vaison la Romaine, France
\and
National Youth Space Center, 59567, Goheung, Jeollanam-do, South Korea
\and
Space Research Institute, Austrian Academy of Sciences, Schmiedlstr. 6, 8042 Graz, Austria
\and
Anunaki Observatory, Calle de los Llanos, 28410 Manzanares el Real, Spain
\and
School of Physical Sciences, The Open University, Milton Keynes, MK7 6AA, UK
\and
Gruppo Astrofili Salese ''Galileo Galilei'', via G. Ferraris 1, 30036 Santa Maria di Sala (VE), Italy
\and
Las Cumbres Observatory, 6740 Cortona Dr., Suite 102, Santa Barbara, CA 93117, USA
\and
Special Astrophysical Observatory, Russian Academy of Sciences, Nizhnij Arkhyz, Russia, 369167
\and
Central Astronomical Observatory at Pulkovo of Russian Academy of Sciences, Pulkovskoje shosse d. 65, St. Petersburg, Russia, 196140
\and
MTA Distinguished Guest Fellow, Konkoly Observatory
\and
NASA Hubble Fellow
\and
Bernoulli Fellow
\and
Gruber Fellow
}

   \date{Received --; accepted --}

  \abstract{Transit events of extrasolar planets offer a wealth of information for planetary characterization. However, for many known targets, the uncertainty of their predicted transit windows prohibits an accurate scheduling of follow-up observations. In this work, we refine the ephemerides of 21 hot Jupiter exoplanets with the largest timing uncertainties. We collected 120 professional and amateur transit light curves of the targets of interest, observed with a range of  telescopes of 0.3m to 2.2m, and analyzed them along with the timing information of the planets discovery papers. In the case of WASP-117b, we measured a timing deviation compared to the known ephemeris of about 3.5 hours, and for HAT-P-29b and HAT-P-31b the deviation amounted to about 2 hours and more. For all targets, the new ephemeris predicts transit timings with uncertainties of less than 6 minutes in the year 2018 and less than 13 minutes until 2025. Thus, our results allow for an accurate scheduling of follow-up observations in the next decade.}

   \keywords{methods: observational --
techniques: photometric -- 
                planets and satellites: fundamental parameters
               }

   \maketitle
%
%-------------------------------------------------------------------

\section{Introduction}

The transit of an extrasolar planet delivers a wealth of information. Time-series photometry of the event allows for the derivation of the orbital period, the orbital inclination, and the planet-star radius ratio \citep{Charbonneau2000,Seager2003}. If the host star is well characterized by high-resolution spectroscopy, a radial velocity curve by a spectroscopic time-series offers the mass of the transiting system \citep[e.g.,][]{Bouchy2005}. This mass in combination with the transit information yields the mean density of the planet.
Transiting systems also provide information on their atmospheric composition through transmission spectroscopy, information on the thermal energy budget through emission spectroscopy, and allow for conclusions on their migration history through the measurement of the misalignment of stellar spin and planetary orbit in accordance with the Rossiter-McLaughlin effect \citep[e.g.,][]{Wakeford2017,Arcangeli2018,Albrecht2012}.

However, any follow-up observation of the transit events, either with photometry, low or high-resolution spectroscopy, or even with polarimetry, relies on reasonably accurate knowledge of the timing of the transit. This knowledge degrades over time because the timing uncertainty increases linearly with the number of transit epochs that have passed since the last observation. The large number of exoplanets discovered per year nowadays makes it more and more difficult to ensure parameter refinement studies for all targets. Therefore, there is a non-negligible number of systems for which the timing uncertainty has reached values of 30~minutes or more. This uncertainty is too high for follow-up observations with space-based or large ground-based telescopes, where observing time is very expensive and good coverage of out-of-transit observations cannot be guaranteed within a limited observing interval. The timing uncertainty can grow so much that the knowledge of when the transit happens is practically lost. Current examples are CoRoT-24b and c \citep{Alonso2014}.

The goal of this work is to refine the ephemeris information of hot Jupiter systems that exhibit large timing uncertainties to ensure the possibility of future follow-up observations. In Section~2, we describe our target selection, in Section~3 we provide details about the observations for this work, and in Section~4 we explain the data analysis. Section~5 provides the results, Section~6 gives a discussion, and in Section~7 we finish the paper by outlining our conclusions.

\section{Target selection}

We compiled a target list with the ephemeris information of hot Jupiter exoplanets from the \textit{NASA Exoplanet Archive} \citep{Akeson2013} and calculated the transit time uncertainty by mid 2018. We constrained our target selection to the planets discovered by ground-based surveys and include all planets named with the prefix WASP, HAT-P, HATS, XO, TrES, KELT, Qatar, and MASCARA. We noticed that several planets discovered by the space mission CoRoT, next to the aforementioned CoRoT-24 b and c, have a timing uncertainty above 30 minutes; examples are CoRoT-16b and 17b with three- and six-hour transit timing uncertainty. However, due to their relative faintness of about fifteenth magnitude in Johnson V, these targets are of limited value for detailed characterization and generally of less interest for follow-up observations. Hence, they are not included here. From the general formula of a planet ephemeris,
\begin{equation}
T_c\,=\,T_0 + n \cdot P \, ,
\end{equation}
with $T_0$ as the timing zero point, $P$ as the orbital period, and $n$ as the number of epochs passed since $T_0$, we calculate the uncertainty of the calculated timing $T_c$ according to the general rules of uncertainty propagation:
\begin{equation}
\Delta T \,=\,\sqrt{\Delta T_0^2 + (n \cdot \Delta P)^2 }\, .
\label{equ_deltaT}
\end{equation}
This equation does not take into account a potential covariance of $T_0$ and $P$, however, for those of our targets with the largest timing uncertainties, $\Delta T$ is strongly dominated by the term $n \cdot \Delta P$ and a potential covariance is of minor importance.

For the 267 targets in our list, the timing uncertainty by August 2018 ranges from 0.3~minutes to 172~minutes with a median of 4~minutes.  
For the 50 objects with the largest timing uncertainties, we verified that the used ephemeris values were still up to date by checking  for each individual target the publications listed in \textit{The Extrasolar Planets Encyclopaedia} \citep{Schneider2011} under ``Related Publications'' and by checking all publications that cited the discovery paper of the target of interest. 

At the time of target selection and data acquisition of this project, HAT-P-25b and HAT-P-38b were ranked position 11 and 8 in Table~\ref{tabtargetsel} with timing uncertainties of 23.7 and 24.6~minutes according to their discovery papers \cite{Quinn2012} and \cite{Sato2012}. In the course of this work, the ephemerides of these planets were improved by \cite{Wang2018} and \cite{Bruno2018}. Both targets are still included here as a consistency check with the new values. For the three hot Jupiters Qatar-3b, -4b, and -5b, the orbital period uncertainty was not provided in their discovery paper \cite{Alsubai2017}. These planets are included here for a new analysis because new results published at the \textit{Exoplanet Transit Database} \citep[ETD,][]{Poddany2010} indicate deviations from the currently known ephemerides for two of the three targets. We finish the ranking in Table~\ref{tabtargetsel} at an uncertainty of 11 minutes. This value is rather arbitrary, but we consider uncertainties smaller than this value to cause only minor problems in the scheduling of transit follow-up observations.

The targets of our sample are very diverse in their planetary and stellar parameters, therefore they are of interest for a broad bandwidth of investigations: from investigations on radius inflation mechanisms of hot Jupiters (Qatar-4b and Kelt-8b have radii larger than 1.5 Jupiter radii) and tidal star-planet interactions (HAT-P-34b, Qatar-3b, Qatar-4b, and Qatar-5b have masses above 3 Jupiter masses), through studies on the formation history of the rare hot Jupiters with planetary companions (HAT-P-44b, HAT-P-45b, HAT-P-46b), to a search for mechanisms that excite the eccentricity of close-in gas giants (HAT-P-31b, HAT-P-34b, WASP-117b). Four host stars among the sample are peculiarly bright with V < 10.5 mag, simplifying any effort for follow-up. WASP-117b is one of the longest-period hot Jupiters and the orbit has been almost unchanged by tidal interaction during its lifetime \citep{Lendl2014}. Because of the large timing uncertainties of our targets, any effort for follow-up observations will greatly benefit from a prior ephemeris refinement.

\begin{table}
\small
\caption{Ranking of hot Jupiter exoplanets according to their timing uncertainties as of August 2018. The targets written in bold face are analyzed in this work.}
\label{tabtargetsel}
\begin{center}
\begin{tabular}{clcr}
\hline
\noalign{\smallskip}
Seq   &  Planet   &  $\Delta T_\mathrm{c}$ (min) & Reference \\
\noalign{\smallskip}
\hline
\noalign{\smallskip}
 1  & \textbf{WASP-73b}  & 171.7   &  \cite{Delrez2014}    \\                 
 2  & \textbf{WASP-117b} & 143.1   &  \cite{Lendl2014}    \\                  
 3  & \textbf{HAT-P-31b} & 106.1   &  \cite{Kipping2011}   \\                 
 4  & \textbf{KELT-8b}   & 103.8    &  \cite{Fulton2015}   \\   
 5  & \textbf{HAT-P-46b} & 40.9    &  \cite{Hartman2014}   \\                    
 6  & \textbf{HAT-P-29b} & 38.8    &  \cite{Buchhave2011}    \\               
 7  & \textbf{HAT-P-45b} & 25.2    &  \cite{Hartman2014}   \\                 
 8  & KELT-10b           & 24.3    &  \cite{Kuhn2016}    \\                   
 9  & \textbf{HAT-P-42b} & 23.7    &  \cite{Boisse2013}    \\                 
10  & \textbf{HAT-P-35b} & 22.9    &  \cite{Bakos2012}    \\                  
11  & WASP-99b           & 21.3    &  \cite{Hellier2014}  \\ 
12  & \textbf{HAT-P-44b} & 16.8    &  \cite{Hartman2014}   \\                 
13  & \textbf{HAT-P-43b} & 15.2    &  \cite{Boisse2013}    \\                 
14  & KELT-15b           & 14.1    &  \cite{Rodriguez2016}   \\               
15  & \textbf{WASP-37b}  & 13.3    &  \cite{Simpson2011}   \\                 
16  & HAT-P-15b          & 13.2    &  \cite{Kovacs2010}   \\ 
17  & \textbf{HAT-P-34b} & 12.3    &  \cite{Bakos2012}   \\                   
18  & \textbf{HAT-P-52b} & 12.2    &  \cite{Hartman2015_52} \\                
19  & \textbf{KELT-3b}   & 12.1    &  \cite{Pepper2013} \\                    
20  & WASP-86/KELT-12b   & 11.6    &  \cite{Faedi2016} \\    
21  & \textbf{WASP-58b}  & 11.1    &  \cite{Hebrard2013}   \\                     
\hline                                                                        
\noalign{\smallskip}                                                          
121  & \textbf{HAT-P-38b} &   4.3  &  \cite{Bruno2018}   \\                   
251  & \textbf{HAT-P-25b} &   0.9  &  \cite{Wang2018}  \\                     
     & \textbf{Qatar-3b} &        &  \cite{Alsubai2017}  \\                   
     & \textbf{Qatar-4b} &        &  \cite{Alsubai2017}  \\                   
     & \textbf{Qatar-5b} &        &  \cite{Alsubai2017}  \\         

\hline
\end{tabular}
\end{center}
\end{table}

\section{Observations and data acquisition}

Transit time-series photometry in the course of this work has been obtained with the 1.2-m STELLA telescope, the 0.8-m Joan Or\'{o} telescope (TJO) of the Montsec Astronomical Observatory, the 2.2-m telescope of the Calar Alto observatory, the 2.15-m Jorge Sahade telescope (JST) of the Observatory Complejo Astron\'omico El Leoncito (CASLEO), the National Youth Space Center (NYSC) 1m telescope, the Chilean-Hungarian Automated Telescope (CHAT), a 1.0-m telescope of the Las Cumbres Observatory (LCO), and Yunnan. 

STELLA and its wide-field imager WiFSIP \citep{Strassmeier2004,Strassmeier2010} observed 19 transits in total in the Sloan r' filter. The original field of view of WiFSIP of 22\,$'\times$ 22\,$'$ was reduced in all observations to 15\,$'\times$ 15\,$'$ to shorten the read-out time. A mild defocus was always applied to spread the PSF to an artificial FWHM of about 3$''$. 

Five transit light curves in Johnson V filter were obtained with the TJO and its main imager MEIA2. The instrument has a field of view of 12.3 x 12.3 arcmin and a resolution of 0.36 arcsec/pixel. All TJO observations employed the Johnson V filter. 

One transit light curve was observed with the Calar Alto 2.2m telescope and its instrument CAFOS in imaging mode. The SITe CCD chip was binned by 2$\,\times\,$2 pixels. Additionally, we applied a read-out window to reduce the read-out time.  
A mild defocus was applied.  

We observed one primary transits of WASP-73b with the JST. The observations were carried out using an R filter and binning 1x1. To increase the size of the field of view to an unvignetted 9 arcmin radius, we employed a focal reducer. 

Five transits were obtained with the NYSC 1m telescope at Deukheung Optical Astronomy Observatory (DOAO) in South Korea with either a FLI PL-16803 CCD camera or a Princeton Instruments SOPHIA-2048B CCD. The telescope was slightly defocused during the observations, and we employed a Cousins R filter.

Two transit light curves were observed with the CHAT, which is a newly commissioned 0.7 telescope at Las Campanas Observatory, Chile, built by members of the HATSouth \citep{Bakos2013} team, and dedicated to the follow-up of transiting exoplanets. A more detailed account of the CHAT facility will be published at a future date (Jord\'an et al., in prep\footnote{https://www.exoplanetscience2.org/sites/default/files/submission-attachments/poster\_aj.pdf}). Data were obtained with a Finger Lakes Microline CCD camera equipped with a back-illuminated CCD and a Sloan i filter. The camera, which has a field of view of $\approx 21\arcmin \times 21\arcmin$, was slightly defocused during the observations.

A light curve of WASP-117b transit was obtained in July 2017 at the LCO, which is a fully robotic network of telescopes \citep{Brown2013}, deployed around the globe in both hemispheres\footnote{For updated information about the network, see: https://lco.global}. We used a 1.0-m telescope of the network at the South African Astrophysical Observatory (SAAO) and the Sinistro camera. The telescope was defocused by 3.0 mm.

One transit light curve was obtained with 1-m telescope of Yunnan Observatories, China, and its 2Kx2K CCD camera using the Cousins R filter. The instrument offers a field of view of 7.2 ,$'\times$7.2\,$'$ and a resolution of 0.2\,$''$/pixel.

We complemented our data sample with a large number of amateur light curves, which we obtained from the ETD\footnote{http://var2.astro.cz/ETD; http://var2.astro.cz/tresca} \citep{Poddany2010}. We selected 85 light curves from this database by visual inspection. While the ETD offers an online fitting routine and provides the derived transit parameter, we downloaded the reduced light curves and re-analyzed them for their timing information homogeneously to the newly obtained light curves with professional telescopes (see Section~\ref{sec_analysis}). A summary of all 120 light curves and their properties is given in Table~\ref{tab_overview}.

Since the data of this paper were obtained by more than 30 different observatories, we did not attempt a homogeneous data reduction. The ETD observers uploaded reduced light curves to the online database and provided some details of the data reduction. The data reduction and light curve extraction of the professional observatories is briefly described in Section~\ref{sec_datared}.

All 120 light curves together with their transit fit (Section~\ref{sec_analysis}) are shown in Figure~\ref{plot_lcs_part1} to \ref{plot_lcs_part5}.

\section{Light-curve analysis}
\label{sec_analysis}

In this work, we analyzed the photometric transit light curves with the publicly available software tool JKTEBOP \citep{Southworth2004,Southworth2008}. Fit parameters were the orbital semi-major axis scaled by the stellar radius $a/R_{\star}$, the orbital inclination $i$, the planet-star radius ratio $k$, the midpoint of the transit $T$, the orbital period $P$, the eccentricity of the orbit $e$, the argument of periastron $\omega$, and coefficients of the detrending function $c_{0,1,2}$.

Many studies suggest that trends in small-telescope transit photometry of 1 to 2~mmag photometric precision is fit by simple detrending functions with very few coefficients \citep[e.g.,][]{Juvan2018,Southworth2016,Mancini2016,Maciejewski2016}. For STELLA/WiFSIP photometry, several studies used the Bayesian Information Criterion to show that first or second order polynomials over time form the best representation of trends or systematics in the light curves \citep[e.g.,][]{Mallonn_H12,Mallonn_H32,Mackebrandt2017}. Therefore, and for the reason that the ETD light curves are lacking the information of external parameters for a more complex detrending, we detrended all light curves of this work consistently by a simple second-order polynomial over time. For the vast majority of targets we present a multiplicity of light curves which warrants a consistency check. 

Crucial for the final derivation of robust uncertainties on the transit timing measurement is a reliable estimation of the photometric uncertainties. We started with the values delivered by the different aperture photometry software tools, which generally include the photon noise of the target, ensemble of comparison stars and background. We ran an initial transit model fit, subtracted the best fit model from the data, and performed a 4-$\sigma$ clipping on the residuals to remove outliers. As a second step, we ran another transit fit and multiplied the photometric uncertainties by a common factor that results in a reduced $\chi^2$ of unity for the fit. Additionally, we calculated the so-called $\beta$ factor, a concept introduced by \cite{Gillon2006} and \cite{Winn2008} to include the contribution of correlated noise in the light curve analysis. It describes the evolution of the standard deviation $\sigma$ of the light-curve residuals when they become binned in comparison to Poisson noise. In the presence of correlated noise, $\sigma$ of the binned residuals is larger by the factor $\beta$ than the binned uncorrelated (white) noise that decreases by the square root of the number of points per bin width. The value of $\beta$ depends on the bin width, we use here the average of ten binning steps from half to twice the duration of ingress. We enlarged the photometric uncertainty finally by this factor $\beta$.

The dates of all light curves were converted to BJD$_\mathrm{TDB}$ \citep{Eastman2010}. All individual transit light curves were now fit with $i$, $a/R_{\star}$, $k$, $P$, $e$, and $\omega$ fixed to literature values. In the case of HAT-P-29\,b, we used the updated parameter values of \cite{Wang2018b}.  
The limb darkening coefficients of the quadratic law were fixed to theoretical values from \cite{Claret2012} and \cite{Claret2013} according to their stellar parameters obtained from the planet discovery papers. ETD light curves taken with a Clear filter were fit with limb darkening coefficients according to Cousins R. The free-to-fit parameters for each individual light curve were $T$ and $c_{0,1,2}$. All individual transit mid-times are summarized in Table~\ref{tab_timing}. 

The estimation of the transit parameter uncertainties was done in JKTEBOP with a Monte Carlo simulation \citep{Southworth2005}, and with a residual-permutation algorithm that takes correlated noise into account \citep{Southworth2008}. The Monte Carlo simulation was run with 5000 steps. As final parameter uncertainties we adopted the larger value of both methods. The uncertainties of the fixed transit parameters were included in the final timing uncertainty by letting them vary during the error estimation within the 1-$\sigma$ ranges of the literature values. 

In the final step, we performed a joint fit of all light curves per target and included $T_0$ with its uncertainty of the previous ephemeris of the discovery paper. Free-to-fit values were $P$ and $T_0$ of a new ephemeris and the detrending coefficients $c_{0,1,2}$ per light curve. The epoch of $T_0$ was chosen to minimize the covariance between $T_0$ and $P$. In the cases of HAT-P-25b and HAT-P-38b, for which refined ephemerides were published in the course of our analysis, we included also the individual transit times that became available (see Table~\ref{tab_timing}). The new ephemerides of the 21 exoplanets of this work are summarized in Table~\ref{tabresult}. In Figure~\ref{plot_OC_part1} to \ref{plot_OC_part3}, we show the individual observed-minus-calculated timing deviations. For HAT-P-29b, \cite{Wang2018b} found the $T_0$ provided in the discovery paper \citep{Buchhave2011} to be affected by an overly small value of the transit duration. Therefore, we used the corrected timings from \cite{Wang2018b} in the joint fit. As a consequence, there is an offset between the displayed previous ephemeris and the corresponding corrected timing values (Figure~\ref{plot_OC_part1}, upper right panel) of the discovery paper.   

We do not attempt a refinement of transit parameters in addition to the ephemerides because a significant fraction of the light curves used here either miss parts of the transit event or do not reach millimag-precision. However, the light curves are available at the Strasbourg astronomical Data Center (CDS) for further use.

\section{Results}

We use the recently published, refined ephemeris values for HAT-P-25b and HAT-P-38b as a cross-check for the values derived in this work. For both planets, the periods deviate only by fractions of the 1-$\sigma$ uncertainties compared to the refined values of \cite{Wang2018} and \cite{Bruno2018}. We were able to increase the precision of the orbital period estimation because we extended the covered time span by one more season. At a late stage in the preparation of this publication, a follow-up study for HAT-P-29b became available \citep{Wang2018b}. We also reached an agreement for the ephemeris within 1-$\sigma$ for this latter work. 

For all individual timings of this study over all targets, we calculate the reduced $\chi^2$ value to be about 1.1. This indicates a reasonable good match between the average deviation from the corresponding linear ephemeris and the measurement uncertainty. A $\chi^2_{\mathrm{red}}$ slightly larger than unity can be caused by starspots in the host-star photosphere that deform the shape of photometric transit curve \citep{Oshagh2013,Holczer2015}.  

For the majority of the targets investigated in this work, the photometric quality of the light curves only allowed for a slight improvement on the precision of $T_0$ compared to the discovery papers. However, including the timing measurement of these publications, our data expand the time interval of observed transit events for all targets significantly. Therefore, the uncertainty of the estimated orbital period $P$ could be lowered by an order of magnitude for all targets. Objects worth emphasizing are WASP-117b, HAT-P-31b, and HAT-P-29b, for which we measured the predicted transit times to be off by more than 2 hours. For WASP-117b, the actual deviation amounted to about 3.5 hours. In the case of HAT-P-35b and WASP-73b, the difference between prediction and measurement was on the order of 1 hour. For HAT-P-29b, the measured timings deviated by 3.7~$\sigma$ from the ephemeris given in the discovery paper. 

\subsection{Comparison to the ETD online fit results}
The Exoplanet Transit Database performs a transit fit to the uploaded light curves \citep{Poddany2010}. The achieved best fit parameters are listed on the webpage and are regularly used in scientific publications \citep[e.g.,][]{Southworth2016,Angerhausen2017,LilloBox2018}. Our re-analyzed timing values of 85 ETD light curves allow for a cross check with these ETD results. We find that on average the absolute timing shows a deviation of only 20\%\ of our 1-$\sigma$ error bars; that is, there is no systematic offset. However, there is a significant scatter of the individual timing differences with a standard deviation of one when expressed in terms of our derived 1-$\sigma$ uncertainties. 
In extreme cases, the deviations between our best-fit values and ETD derived parameters reached 4\,$\sigma$. We find that the ETD error bars are on average smaller by a factor of 1.7 than the corresponding values derived by the standard procedures used in this work. The ETD adopts the parameter uncertainties from a Levenberg–Marquardt optimization algorithm \citep{Poddany2010}, which is believed to be unreliable in the presence of parameter correlations \citep[see][and references therein]{Southworth2008}.  Therefore, we recommend the re-analysis of ETD light curves instead of the usage of the transit parameters obtained by the online fitting tool.

\begin{table*}
\small
\caption{Refined ephemerides resulting from this work.}
\label{tabresult}
\begin{center}
\begin{tabular}{lp{1.8cm}p{0.03cm}p{1.5cm}p{1.2cm}p{0.03cm}p{1.3cm}r}
\hline
\noalign{\smallskip}
Planet   &  \multicolumn{3}{c}{T$_0$ [BJD\_{TDB}]} & \multicolumn{3}{c}{$P$ (days)} & Reference \\
\noalign{\smallskip}
\hline
\noalign{\smallskip}
HAT-P-25b   &  2455176.85173 &$\pm$& 0.00047   &  3.652836   &$\pm$& 0.000019   &  \cite{Quinn2012}  \\
            &  2456418.80996 &$\pm$& 0.00025   &  3.65281572 &$\pm$& 0.00000095 & \cite{Wang2018}  \\
            &  2457006.91299 &$\pm$& 0.00021   &  3.65281591 &$\pm$& 0.00000067 & this work \\
\hline
\noalign{\smallskip}
HAT-P-29b   &  2455197.57617 &$\pm$& 0.00181   &  5.723186  &$\pm$& 0.000049  &  \cite{Buchhave2011}   \\
            &  2456170.5494  &$\pm$& 0.0015    &  5.723390  &$\pm$& 0.000013  & \cite{Wang2018b} \\
            &  2457092.00345 &$\pm$& 0.00128   &  5.7233773 &$\pm$& 0.0000072 & this work \\
\hline
\noalign{\smallskip}
HAT-P-31b   &  2454320.8866 &$\pm$& 0.0051   &   5.005425  &$\pm$&  0.000091   &  \cite{Kipping2011} \\
            &  2458169.9410 &$\pm$& 0.0017   &   5.0052724  &$\pm$& 0.0000063     &  this work  \\
\hline
\noalign{\smallskip}
HAT-P-34b   &  2455431.59706 &$\pm$& 0.00055   &   5.452654  &$\pm$& 0.000016   &  \cite{Bakos2012} \\
            &  2456462.14718 &$\pm$& 0.00053   &   5.4526470 &$\pm$&  0.0000031  &  this work  \\
\hline
\noalign{\smallskip}
HAT-P-35b   &  2455578.66158 &$\pm$& 0.00050   &  3.646706  &$\pm$& 0.000021  &  \cite{Bakos2012} \\
            &  2456836.75811 &$\pm$& 0.00041   &  3.6466566 &$\pm$& 0.0000012  &  this work  \\
\hline
\noalign{\smallskip}
HAT-P-38b   &  2455863.12034 &$\pm$& 0.00035   &  4.640382  &$\pm$& 0.000032   & \cite{Sato2012} \\
            &                &     &           &  4.6403294 &$\pm$& 0.0000055  &  \cite{Bruno2018} \\
            &  2457491.87585 &$\pm$& 0.00009   &  4.6403293 &$\pm$& 0.0000017  & this work \\
\hline
\noalign{\smallskip}
HAT-P-42b   &  2455952.52683 &$\pm$& 0.00077   &  4.641876  &$\pm$& 0.000032  &  \cite{Boisse2013} \\
            &  2456036.07987 &$\pm$& 0.00077   &  4.6418381 &$\pm$&  0.0000080  &  this work \\
\hline
\noalign{\smallskip}
HAT-P-43b   &  2455997.37182 &$\pm$& 0.00032   &  3.332687  &$\pm$& 0.000015  &  \cite{Boisse2013} \\
            &  2456147.34248 &$\pm$& 0.00030   &  3.3326830 &$\pm$& 0.0000019  &  this work \\
\hline
\noalign{\smallskip}
HAT-P-44b   &  2455696.93772 &$\pm$& 0.00024  &  4.301219  &$\pm$& 0.000019  &  \cite{Hartman2014}  \\
            &  2456204.47794 &$\pm$& 0.00019  &  4.3011886 &$\pm$& 0.0000010 &  this work  \\
\hline
\noalign{\smallskip}
HAT-P-45b   &  2455729.98689 &$\pm$& 0.00041  &  3.128992  &$\pm$& 0.000021  &  \cite{Hartman2014}  \\
            &  2456502.84809 &$\pm$& 0.00033  &  3.1289923 &$\pm$& 0.0000014  &  this work  \\
\hline
\noalign{\smallskip}
HAT-P-46b   &  2455701.33723 &$\pm$& 0.00047   &  4.463129  &$\pm$& 0.000048  &  \cite{Hartman2014}  \\
            &  2455969.12547 &$\pm$& 0.00044   &  4.4631365 &$\pm$& 0.0000050  &  this work  \\
\hline
\noalign{\smallskip}
HAT-P-52b   &  2455852.10403 &$\pm$& 0.00041   &   2.7535953 &$\pm$& 0.0000094  &  \cite{Hartman2015_52} \\
            &  2456645.13981 &$\pm$& 0.00032   &   2.7535965 &$\pm$& 0.0000011  &  this work  \\
\hline
\noalign{\smallskip}
KELT-3b     &  2456034.29537 &$\pm$& 0.00038   &   2.703390  &$\pm$& 0.000010  &  \cite{Pepper2013} \\
            &  2456269.48987 &$\pm$& 0.00029   &   2.7033850 &$\pm$& 0.0000018 &  this work \\
\hline
\noalign{\smallskip}
KELT-8b     &  2456883.4803  &$\pm$&  0.0007    &   3.24406   &$\pm$& 0.00016  &  \cite{Fulton2015} \\
            &  2457396.04496 &$\pm$&  0.00055   &   3.2440796 &$\pm$& 0.0000048  &  this work \\
\hline
\noalign{\smallskip}
Qatar-3b    &  2457302.45300 &$\pm$& 0.00010  &  2.5079204 &     &        & \cite{Alsubai2017} \\
            &  2457312.48458 &$\pm$& 0.00010  &  2.5078952 &$\pm$& 0.0000032  &  this work  \\
\hline
\noalign{\smallskip}
Qatar-4b    &  2457637.77361 &$\pm$& 0.00046  &  1.8053564 &     &        & \cite{Alsubai2017} \\
            &  2457872.47170 &$\pm$& 0.00046  &  1.8053704 &$\pm$& 0.0000042  &  this work  \\
\hline
\noalign{\smallskip}
Qatar-5b    &  2457336.75824 &$\pm$& 0.00010 &  2.8792319 &     &        & \cite{Alsubai2017} \\
            &  2457362.67203 &$\pm$& 0.00009 &  2.8793105 &$\pm$& 0.0000025  &  this work  \\
\hline
\noalign{\smallskip}
WASP-37b    &  2455338.6196  &$\pm$& 0.0006    &   3.577469  &$\pm$& 0.000011  &  \cite{Simpson2011} \\
            &  2456393.97698 &$\pm$& 0.00052   &   3.5774807 &$\pm$& 0.0000019  &  this work  \\
\hline
\noalign{\smallskip}
WASP-58b    &  2455183.9342  &$\pm$& 0.0010     &   5.017180  &$\pm$& 0.000011  &  \cite{Hebrard2013} \\
            &  2457261.05970 &$\pm$& 0.00062    &   5.0172131 &$\pm$& 0.0000026  &  this work \\
\hline
\noalign{\smallskip}
WASP-73b    &  2456128.7063 &$\pm$& 0.0011     &   4.08722  &$\pm$&  0.00022 &  \cite{Delrez2014} \\
            &  2456365.7688 &$\pm$&  0.0011   &   4.0872856 &$\pm$&  0.0000087  &  this work \\
\hline
\noalign{\smallskip}
WASP-117b   &  2456533.82404 &$\pm$& 0.00095   &  10.02165  &$\pm$& 0.00055  &  \cite{Lendl2014} \\
            &  2457355.51373 &$\pm$& 0.00055   &  10.020607 &$\pm$& 0.000011 &  this work  \\
\hline
\noalign{\smallskip}
\hline
\end{tabular}
\end{center}
\end{table*}

\section{Discussion}
We use our sample of 21 newly determined ephemerides to check statistically if the differences between old and new ephmerides are in general agreement to the previous ephemeris uncertainties. When we express the measured-to-predicted timing deviation of all our 21 targets in units of their previously known timing uncertainties, we find a standard deviation for all targets of about 1.4~one-sigma uncertainties; that is, larger than unity. This indicates a trend of slightly underestimated uncertainties of the ephemerides. There may be various reasons for this depending on individual targets; for example, underestimated systematics in the data, systematic effects on the host star, like stellar activity, or transit timing variations (TTV).  

We consider it to be possible that significant timing deviations from the predicted values originate from TTVs. Such variation could be caused by gravitational interactions of the observed hot Jupiter with unknown planetary companions \citep[][and references therein]{vonEssen2018}. Hot Jupiter planets are known to mostly orbit their host star alone \citep{Steffen2012}. However in recent years a few exceptions to this general rule have been found, such as the planetary system of WASP-47 with one hot Jupiter accompanied by an interior and an exterior sub-Neptune \citep{Becker2015, Neveu2016}. Using all 3.5 years of Kepler spacecraft data, \cite{Huang2016} showed that these exceptions are very rare. For two of these systems, HAT-P-13 and WASP-47, the literature provides constraints on the TTV amplitude caused by the companions. In the case of HAT-P-13b, \cite{Fulton2011} ruled out periodic TTV of an amplitude larger than 144~s, while for WASP-47b, \cite{Becker2015} measured a TTV amplitude of 38~s. The planetary systems of the hot Jupiters WASP-53b and WASP-81b are uncommon in that they also each harbor an eccentric brown dwarf within a few astronomical units of the host star. Predicted TTVs of these hot Jupiters are below 1 minute \citep{Triaud2017}.

Among our target list, there are three systems with RV candidate signals of Jupiter-mass companions within 1 AU, HAT-P-44, HAT-P-45, and HAT-P-46 \citep{Hartman2014}. Our newly derived ephemerides of HAT-P-45b and HAT-P-46b are in very good agreement with the ones previously published by \cite{Hartman2014}. The individual data points from different seasons show no significant deviation from the linear ephemerides, and therefore we find no indications for significant effects of TTVs. On the other hand, the eight individual measurements of HAT-P-44b show a rather large reduced $\chi^2$ value of 2.3. Nevertheless, we find no TTV periodicity at the planet companion period of about 220~days. The newly derived value of the period deviates by about 25~minutes from the discovery paper. To compute an order of magnitude of the amplitude of TTVs of planet b caused by the outer perturber, we made use of \textit{TTVFast} \citep{Deck2014}. Here we assumed coplanar orbits, a circular orbit for the perturber, and the masses and periods from \cite{Hartman2014}. The derived TTV amplitude for planet HAT-P-44b is about 6~s, which is extremely challenging to measure for ground-based observatories. With increasing mutual inclination, the mass of the outer perturber would also increase due to the degeneracy of $M$ with sin$i$, and so would the TTV amplitude \citep{Payne2011}. However, we consider it to be likely that TTVs only have a marginal effect on the deviation of 25~minutes. It is more probable that this deviation is caused by the limited precision of the previous ephemeris, since it amounts to only 1.6\,$\sigma$, which we do not consider as significant.

An indicator for the potential existence of an outer perturber causing TTVs might also be a nonzero eccentricity of the hot Jupiter. Among our target list, there are four targets with an $e$ significantly different from zero: WASP-117b, HAT-P-29b, HAT-P-31b, and HAT-P-34b. For all four targets, both the number of individual transit epochs and their individual precision is too low to allow for conclusions on TTVs with approximately  one-minute amplitudes. We increase the precision of the period determination by an order of magnitude, and therefore our new ephemerides form the best available basis for future follow-up studies. 

The target HAT-P-29b shows the most significant deviation of measured-to-predicted transit timings with a significance level of 3.7, a deviation very recently also described by \cite{Wang2018b}. In this particular case, a likely explanation is the sensitivity of the used partial transit light curves to an overly small value of the transit duration derived in the discovery paper. For more details, we point the reader to the discussion supplied in \cite{Wang2018b}.

\section{Conclusion}

We have refined the ephemerides of 21 exoplanets which previously had the largest timing uncertainties of ground-based detected hot Jupiters. We made use of a total of 120 transit light curves: 35 obtained from professional observatories and 85 from amateur observers. The bulk of our data might be considered as data of only moderate photometric quality, since more than half of our light curves have a point-to-point scatter larger than 3~mmag, or lack the ingress or egress part of the transit. However, the present work is a valuable example of where light curves of small-sized telescopes can still play a crucial role in modern science. All data were analyzed homogeneously, and resulted in an increased precision in the estimations of the orbital periods by one order of magnitude when combined with the transit-timing information of the discovery papers. Previous to our work, the timing uncertainty of the 21 analyzed objects ranged from 11 to 171~minutes. We were able to lower this to values ranging from 1 to 6~minutes, and thus to ensure a reasonable scheduling of follow-up studies at least until the year 2025, when the timing uncertainties will still be below 12~minutes for all our targets. Our new ephemerides might be affected to a certain extent by stellar activity and TTVs caused by unknown companions. The former constitutes a form of correlated noise in the data and is accounted for in the error estimation, while the latter is less likely because additional companions to hot Jupiters are extremely rare and cause TTV of low amplitude. In any case, we emphasize that even in the cases of ephemerides affected by astrophysical disturbances, our new ephemerides present the best available basis for future follow-up studies.
Currently, the ground-based detected hot Jupiter with the largest timing uncertainty is KELT-10b with $\Delta T_\mathrm{c} \approx 24$~min (by August 2018). We note that especially due to the enormous effort of the observers of the Exoplanet Transit Database, there is currently no hot Jupiter known in the northern hemisphere discovered by ground-based surveys with a timing uncertainty larger than 14~minutes.

\begin{acknowledgements}
E.H. acknowledges support by the Spanish Ministry of Economy and Competitiveness (MINECO) and the Fondo Europeo de Desarrollo Regional (FEDER) through grant ESP2016-80435-C2-1-R, as well as the support of the Generalitat de Catalunya/CERCA program. D. D. acknowledges support provided by NASA through Hubble Fellowship grant HST-HF2-51372.001-A awarded by the Space Telescope Science Institute, which
is operated by the Association of Universities for Research in Astronomy, Inc., for NASA, under contract NAS5-26555. E.S. acknowledges support by the Russian Science Foundation grant No. 14-50-00043 for conducting international photometric observing campaign. I.S. acknowledges support by the Russian Foundation for Basic Research (project No. 17-02-00542). R.B.\ acknowledges support from FONDECYT Post-doctoral Fellowship Project No. 3180246. M.L. acknowledges support from the Austrian Research Promotion Agency (FFG) under project 859724 "GRAPPA". S.-H.G. acknowledges the financial support from the National Natural Science Foundation of China (No.U1531121). L.C. recognizes funding from DLR - 50OR1804. Funding for the Stellar Astrophysics Centre is provided by The Danish National Research Foundation (Grant agreement no.: DNRF106). We thank the staff at the various participating observatories STELLA, CASLEO, Calar Alto, TJO, and others for their support. Partly based on data obtained with the STELLA robotic telescopes in Tenerife, an AIP facility jointly operated by AIP and IAC, on data acquired at Complejo Astron\'{o}mico El Leoncito, operated under agreement between the Consejo Nacional de Investigaciones Cient\'{\i}ficas y T\'{e}cnicas de la Rep\'{u}blica Argentina and the National Universities of La Plata, C\'{o}rdoba and San Juan, on observations collected at the Centro Astron\'{o}mico Hispano Alem\'{a}n (CAHA) at Calar Alto, operated jointly by the Max-Planck Institut f\"{u}r Astronomie and the Instituto de Astrof\'{\i}sica de Andaluc\'{\i}a (CSIC), and on data obtained with the Joan Oró Telescope (TJO) of the Montsec Astronomical Observatory (OAdM), owned by the Generalitat de Catalunya and operated by the Institute for Space Studies of Catalonia (IEEC). This research has made use of the SIMBAD data base and VizieR catalog access tool, operated at CDS, Strasbourg, France, and of the NASA Astrophysics Data System (ADS).
\end{acknowledgements}

\bibliographystyle{aa} % style aa.bst
\bibliography{ERPSE_bib.bib} % your references Yourfile.bib
%\bibliography{test} % your references Yourfile.bib

\begin{appendix} %First appendix
\section{Data reduction}
\label{sec_datared}

The data reduction of the STELLA and CalarAlto data was done with a customized pipeline already used for previous transit light curve analyses \citep{Mallonn_H12,Mallonn_H32,Mackebrandt2017}. Aperture photometry was done with the publicly available software SExtractor \citep{Bertin1996}. We selected the aperture size that minimized the scatter in the light curve residuals after subtraction of an initial transit model of literature transit parameters including a second-order polynomial over time for detrending. Using the same criterion of minimization of photometric scatter, our data reduction pipeline also automatically chose the best selection of comparison stars for differential photometry. 

The TJO imaging frames were reduced using the ICAT reduction pipeline at the TJO \citep{Colome2006} and aperture photometry was extracted using AstroImageJ.

The data reduction of the JST light curves was carried out using DIP2OL, which structure and use is fully explained in \cite{vonEssen2018}. Briefly, the first part of the pipeline is IRAF-based; it carries out the pre-reduction, the extraction of stellar fluxes, and the computation of photometric uncertainties in an automatized way. The second part of the pipeline is a python program that minimizes the scatter of the differential light curves by the most adequate combination of reference stars for the differential light curve, and an optimization of the aperture and the radius size where the sky counts are determined.

The four imaging time-series from the NYSC 1-m telescope were reduced by the IRAF \textit{ccdred} package and aperture photometry was performed with SExtractor. Differential photometry was obtained by the usage of an ensemble of comparison stars.

The CHAT photometric images were reduced using a dedicated pipeline descendant of a pipeline created to obtain photometry using the Las Cumbres 1-m telescopes \citep[][Espinoza et al., in prep]{Shporer2017}, which was also used for the LCO imaging time-series analyzed in this work.

The data reduction of the Yunnan observatory data were reduced using the IRAF package, and systematic errors were removed from the resulting photometric data according to the procedures in \cite{Wang2014}.
\section{Tables and figures}

%\begin{center}
\begin{figure*}
\includegraphics[width=16.5cm]{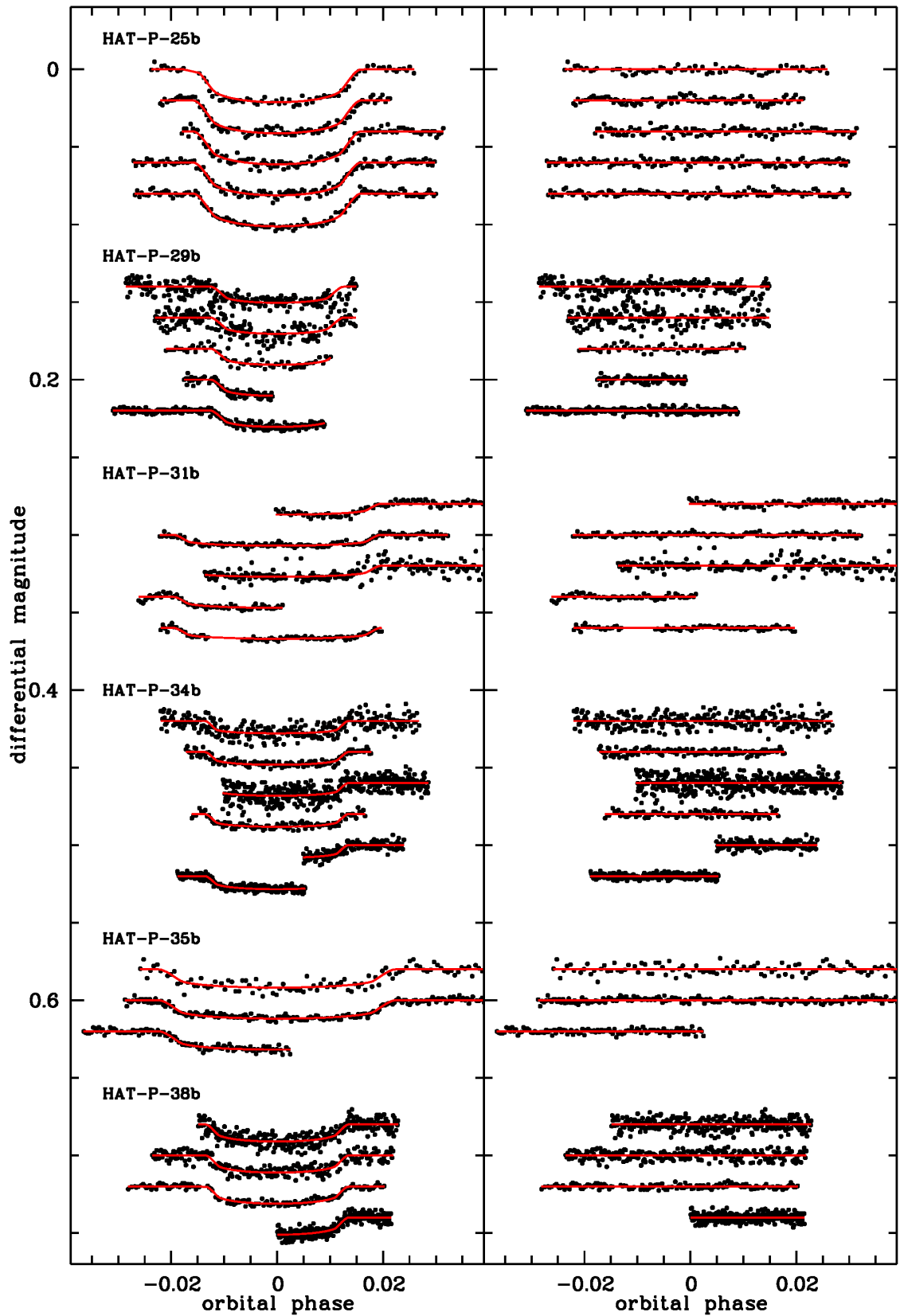}
\caption{Detrended transit light curves in the same order as in Table~\ref{tab_overview}. Curves after the first are displaced vertically for clarity. Residuals from the fits are displayed in the right panel with the same vertical offset.}
\label{plot_lcs_part1}
\end{figure*}
%\end{center}

%\begin{center}
\begin{figure*}
\includegraphics[width=16.5cm]{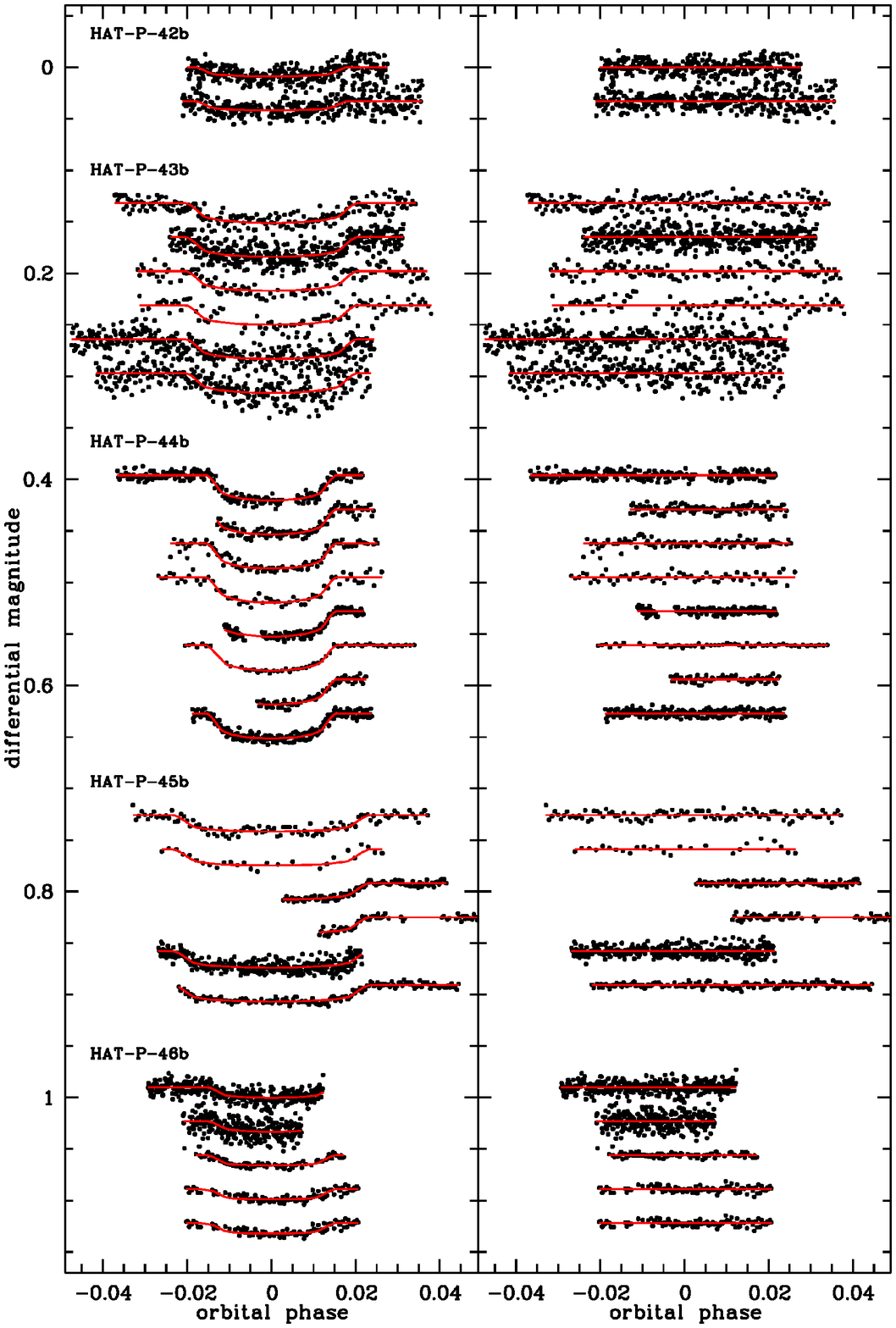}
\caption{Continuation of Figure~\ref{plot_lcs_part1}. }
\label{plot_lcs_part2}
\end{figure*}
%\end{center}

%\begin{center}
\begin{figure*}
\includegraphics[width=16.5cm]{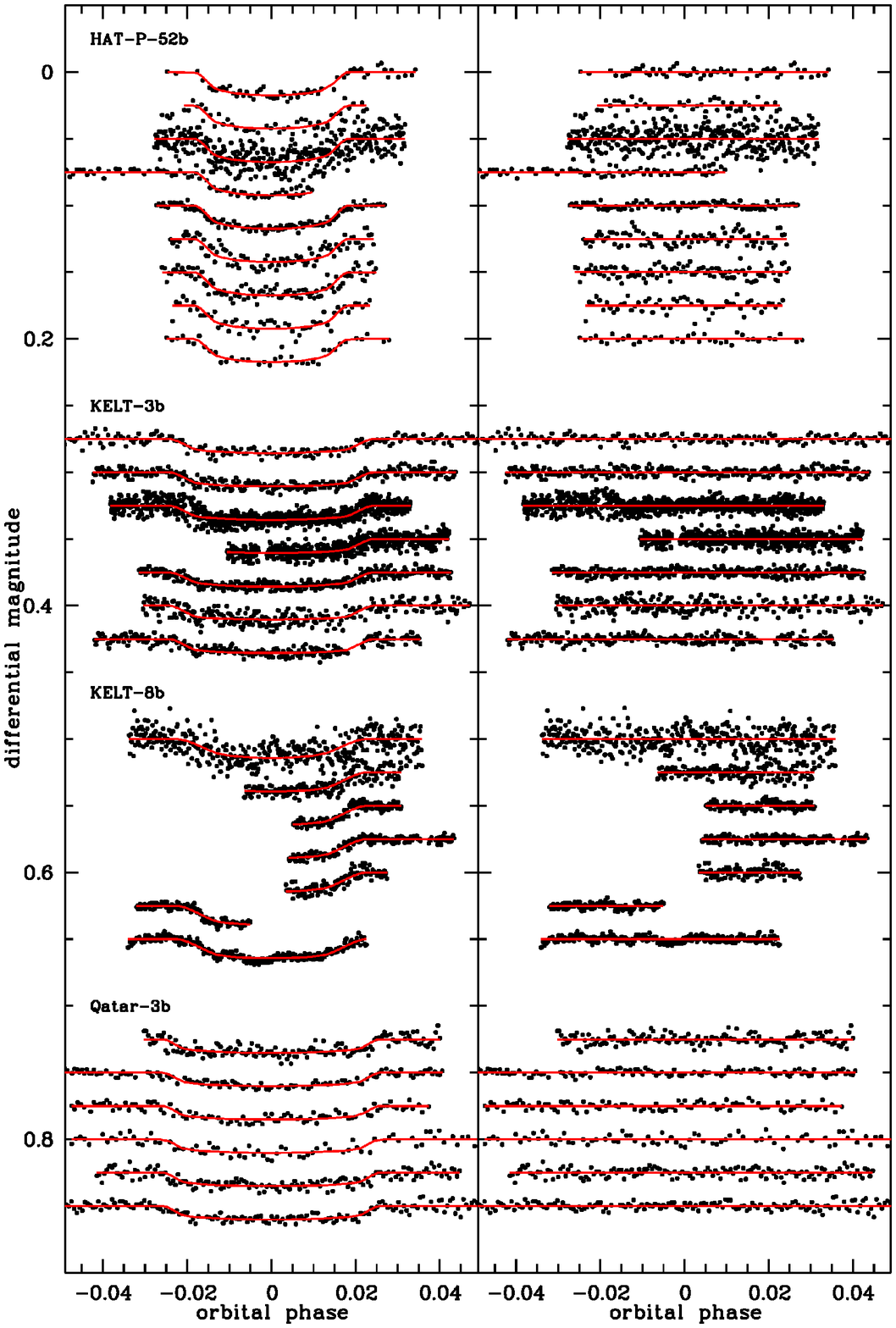}
\caption{Continuation of Figure~\ref{plot_lcs_part1}.}
\label{plot_lcs_part3}
\end{figure*}
%\end{center}

%\begin{center}
\begin{figure*}
\includegraphics[width=16.5cm]{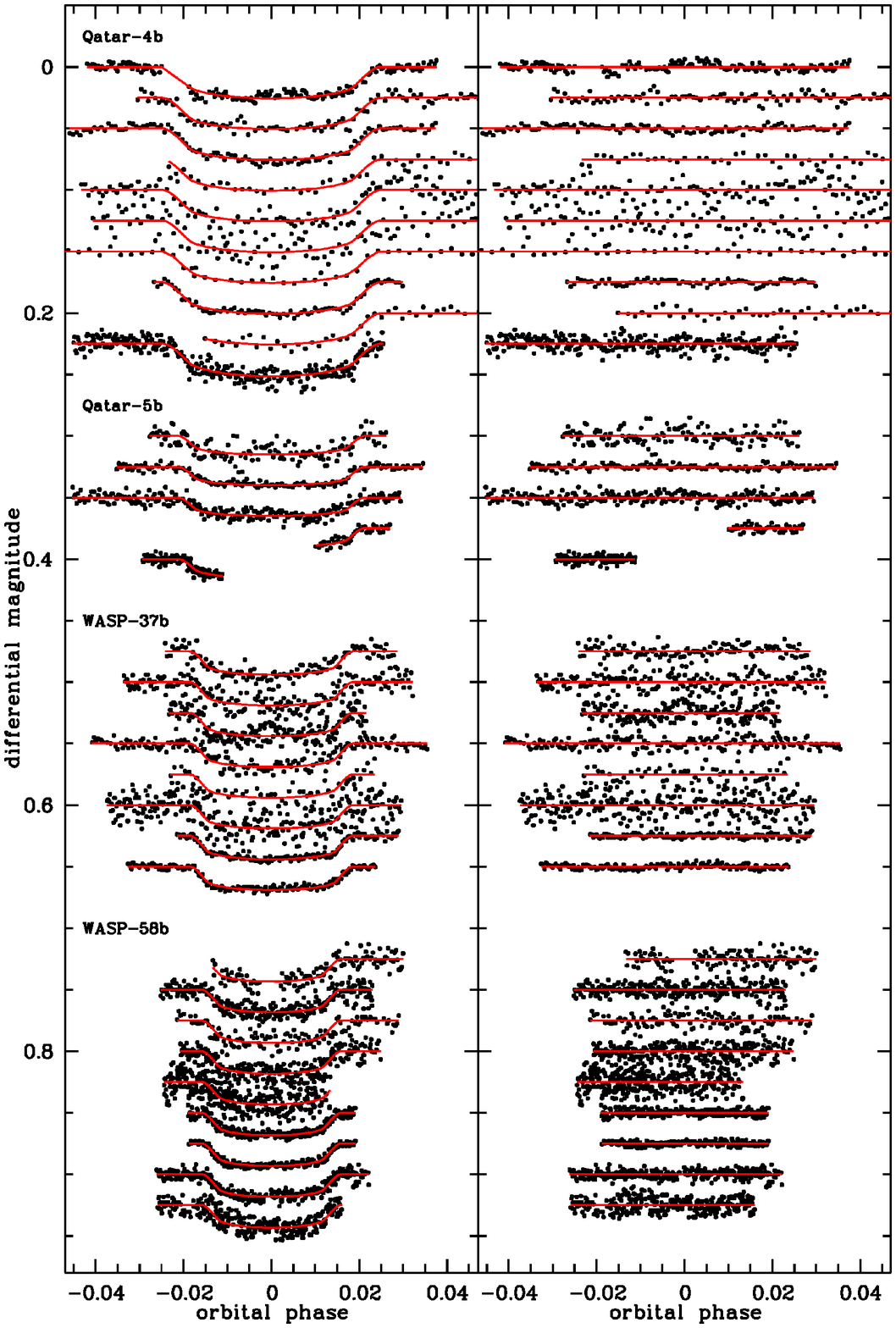}
\caption{Continuation of Figure~\ref{plot_lcs_part1}. }
\label{plot_lcs_part4}
\end{figure*}
%\end{center}

%\begin{center}
\begin{figure*}
\includegraphics[width=16.5cm]{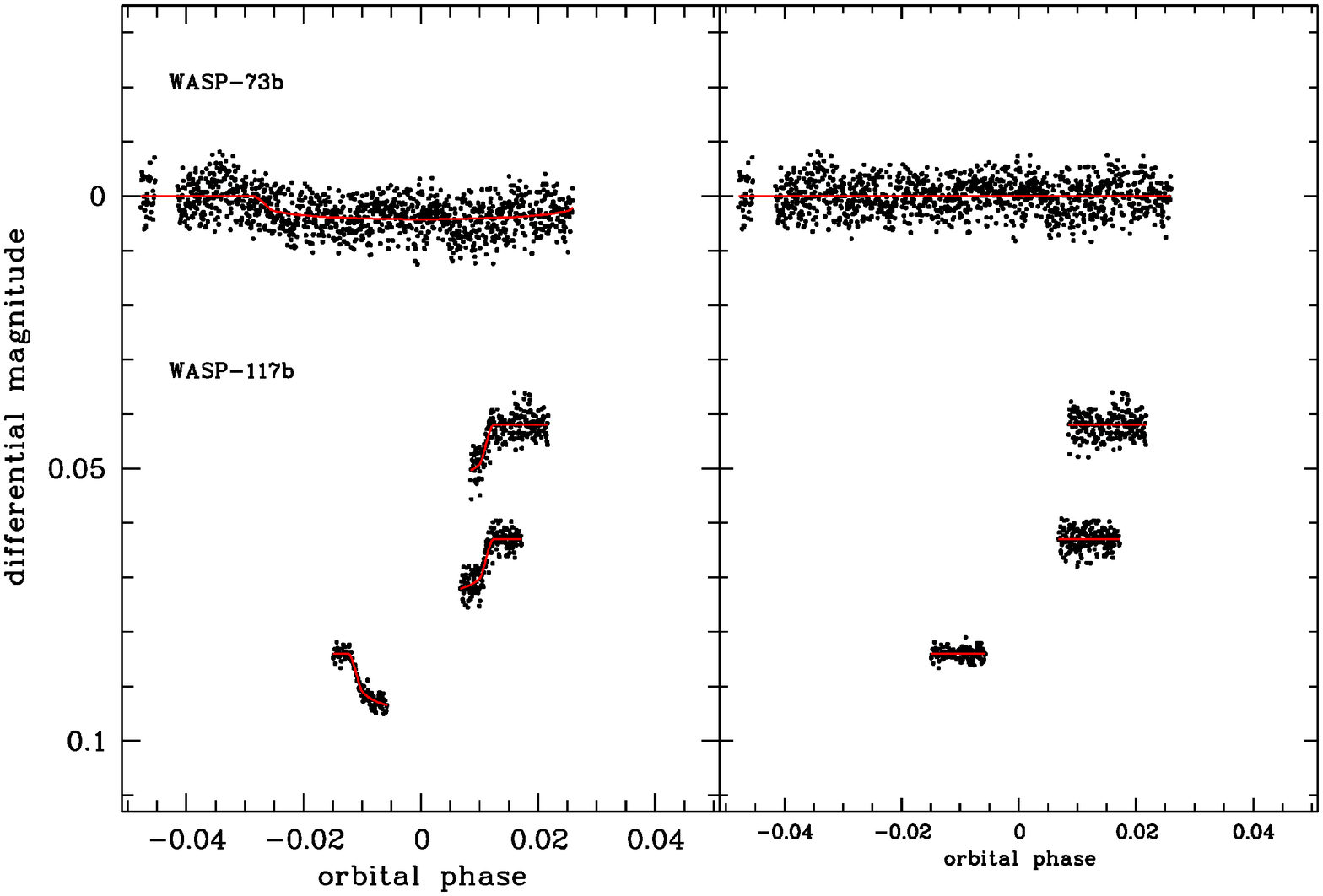}
\caption{Continuation of Figure~\ref{plot_lcs_part1}.}
\label{plot_lcs_part5}
\end{figure*}
%\end{center}

%\begin{center}
\begin{figure*}
\includegraphics[width=16.1cm]{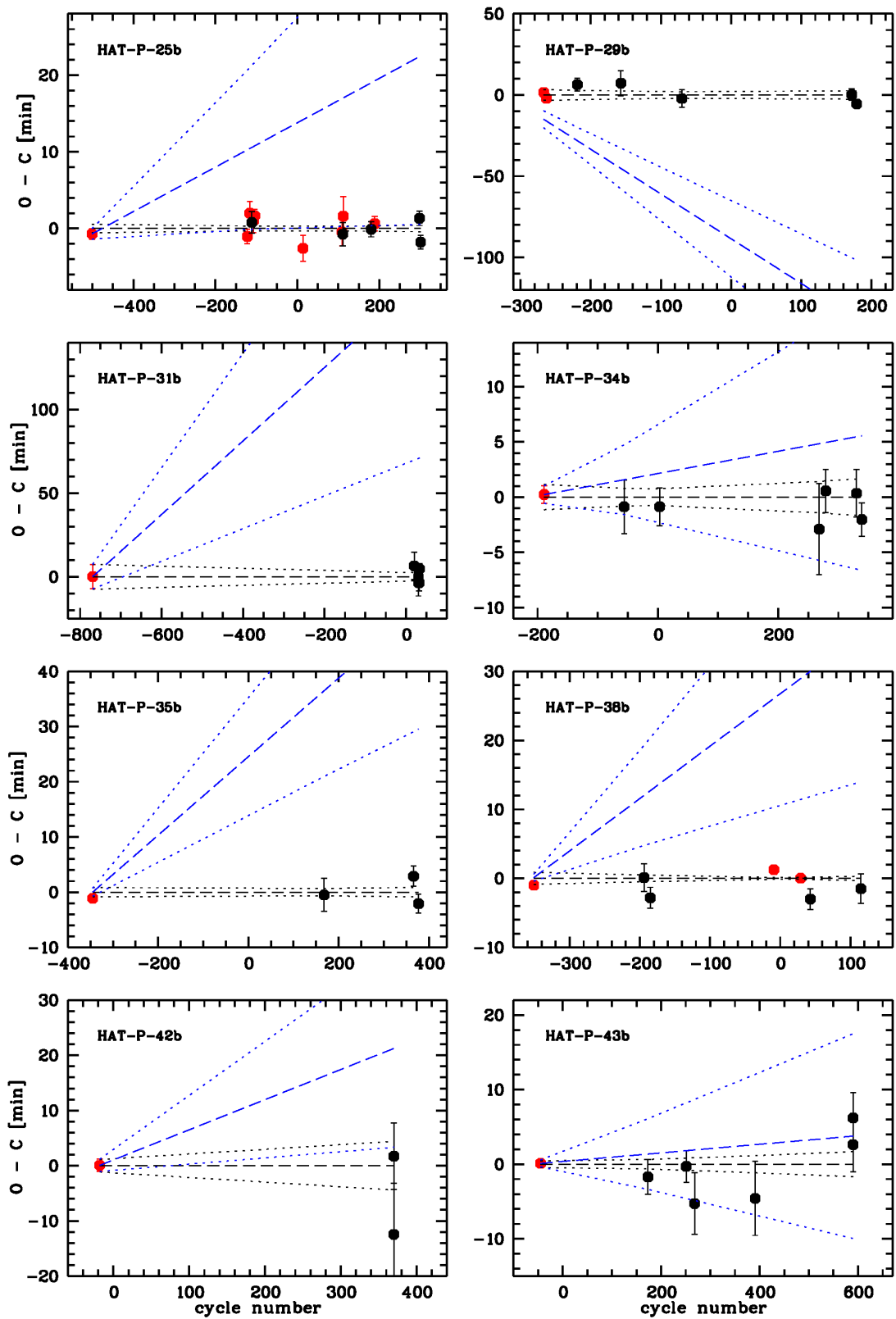}
\caption{Observed minus calculated mid-transit times for HAT-P-25b, HAT-P-29b, HAT-P-31b, HAT-P-34b, HAT-P-35b, HAT-P-38b,  HAT-P-42b, and HAT-P-43b. Measurements of this work in black, literature values included in our calculation in red. A black dashed line denotes the new ephemeris of this work with associated uncertainties in dotted lines. For comparison, the previous ephemeris of the discovery paper in blue. The offset between previous ephemeris and literature value for HAT-P-29b is caused by a timing offset corrected in \cite{Wang2018b}; see text for details. }
\label{plot_OC_part1}
\end{figure*}
%\end{center}

%\begin{center}
\begin{figure*}
\includegraphics[width=16.5cm]{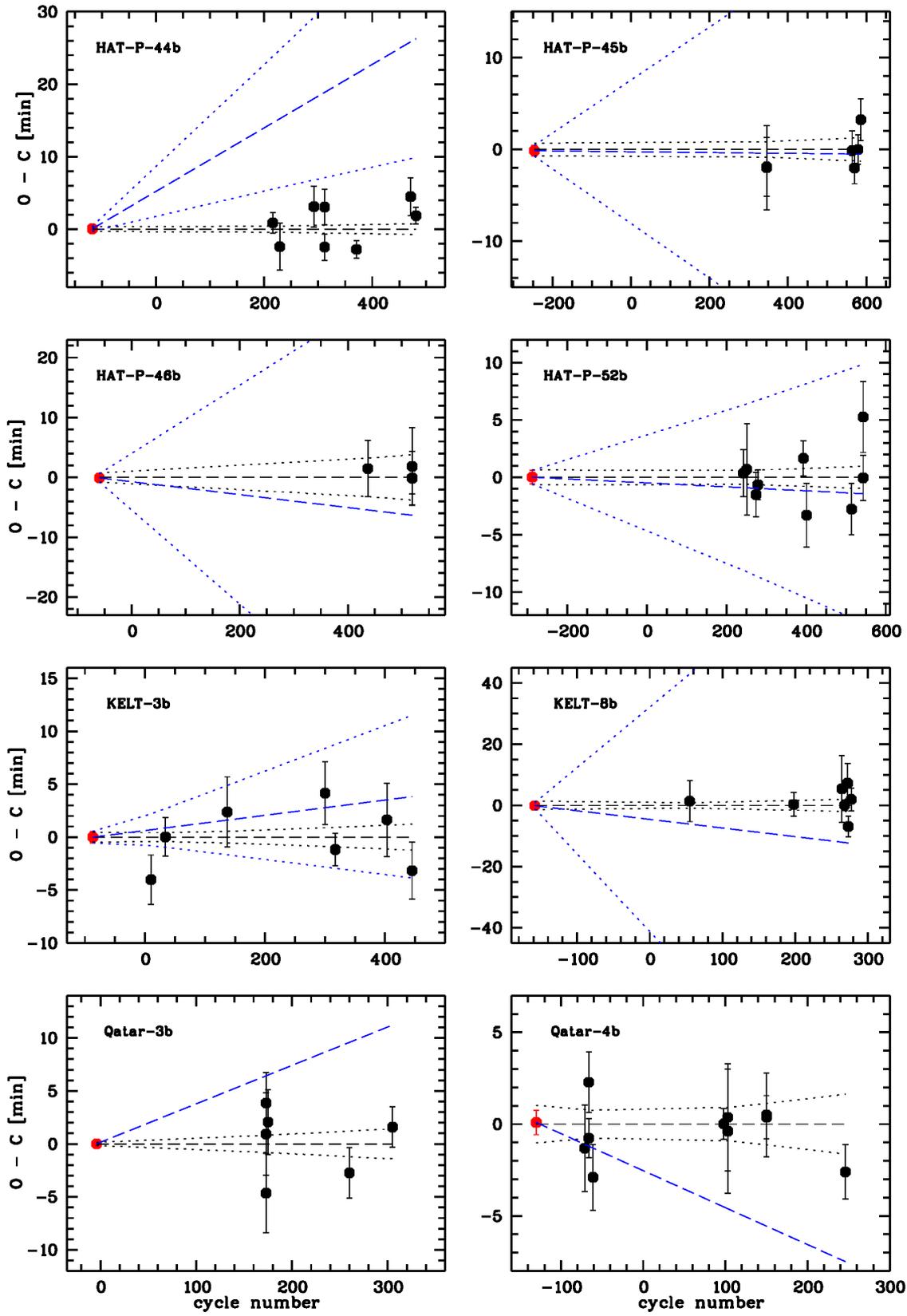}
\caption{Continuation of Figure~\ref{plot_OC_part1} for HAT-P-44b, HAT-P-45b, HAT-P-46b, HAT-P-52b, KELT-3b, KELT-8b, Qatar-3b, and Qatar-4b. We note that Qatar-3b and Qatar-4b lack an ephemeris uncertainty in their discovery paper.}
\label{plot_OC_part2}
\end{figure*}
%\end{center}

%\begin{center}
\begin{figure*}
\includegraphics[width=16.5cm]{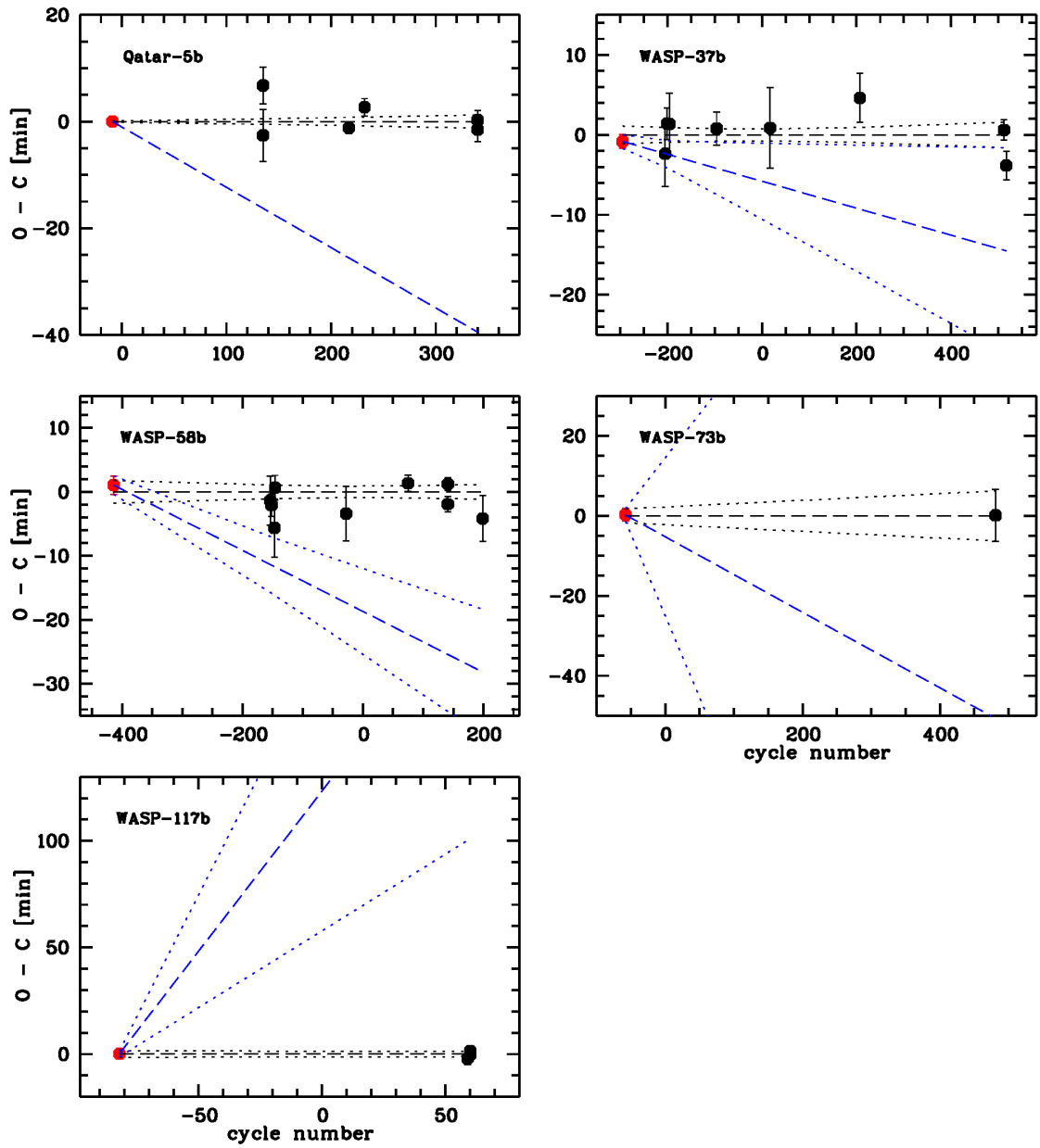}
\caption{Continuation of Figure~\ref{plot_OC_part1} for Qatar-5b, WASP-37b, WASP-58b, WASP-73b, and WASP-117b. We note that Qatar-5b lacks an ephemeris uncertainty in its discovery paper.}
\label{plot_OC_part3}
\end{figure*}
%\end{center}

\longtab[1]{
\begin{longtable}{llllccc}
\caption{\label{tab_overview}Overview about the transit observations of the investigated planets.}\\
\hline
\hline
\noalign{\smallskip}
Planet  &  Date        & Telescope &  Filter  & $N_{\mathrm{data}}$ &  rms (mmag) &  $\beta$  \\
\hline
\noalign{\smallskip}
\endfirsthead
\caption{Continued.} \\
\hline
\noalign{\smallskip}
Planet  &  Date        & Telescope &  Filter  & $N_{\mathrm{data}}$ &  rms (mmag) &  $\beta$  \\
\hline
\noalign{\smallskip}
\endhead
\hline
\endfoot
\hline
\endlastfoot
HAT-P-25b &  2013, Nov 4  &  ETD, M. Salisbury    &  Clear  &  72   &  1.9  &  1.41  \\
          &  2016, Jan 24 &  ETD, M. Bretton      &  Clear  &  102  &  2.0  &  2.01  \\
          &  2016, Oct 2  &  ETD, V.-P. Hentunen  &  Clear  &  119  &  2.0  &  1.21  \\
          &  2017, Dec 7  &  STELLA               &  r'     &  135  &  1.7  &  1.13  \\
          &  2017, Dec 18 &  STELLA               &  r'     &  133  &  1.5  &  1.31  \\
\hline
\noalign{\smallskip}
HAT-P-29b & 2011, Oct 3   &  ETD, M. Vanhuysse    &  R      &  251  &   3.4  &  1.72  \\
          & 2012, Sep 22  &  ETD, J. Trnka        &  Clear  &  204  &   4.6  &  1.51  \\
          & 2014, Feb 2   &  ETD, M. Salisbury    &  R      &  61   &   1.8  &  1.22  \\
          & 2017, Nov 18  &  STELLA               &  r'     &  88   &   1.6  &  1.48  \\
          & 2017, Dec 28  &  STELLA               &  r'     &  239  &   1.3  &  1.00  \\
\hline
\noalign{\smallskip}          
HAT-P-31b & 2018, May 31  &  NYSC 1m              &  R      &  111  &   1.6  &  2.03  \\
          & 2018, Jul 20  &  NYSC 1m              &  R      &  129  &   1.1  &  1.28  \\
          & 2018, Jul 20  &  Yunnan               &  R      &  253  &   3.3  &  1.06  \\    
          & 2018, Jul 30  &  NYSC 1m              &  R      &   65  &   1.4  &  1.46  \\
          & 2018, Aug 04  &  NYSC 1m              &  R      &   84  &   1.1  &  1.38  \\        
\hline
\noalign{\smallskip}
HAT-P-34b & 2012, Aug 17  &  ETD, S. Shadick      &  I      &  279  &  4.1  & 1.29   \\
          & 2013, Jul 4   &  ETD, J. Gonzalez     &  Clear  &  126  &  1.5  & 1.01   \\
          & 2017, Jun 18  &  ETD, F. Scaggiante   &  R      &  437  &  4.3  & 1.14   \\
          & 2017, Aug 17  &  ETD, F. Campos       &  R      &   97  &  1.9  & 1.60   \\
          & 2018, May 23  &  STELLA               &  r'     &  135  &  2.3  & 1.00   \\
          & 2018, Jul 11  &  TJO                  &  V      &  317  &  1.3  & 1.75   \\
\hline
\noalign{\smallskip}
HAT-P-35b & 2016, Mar 1   &  ETD, D. Molina       &  Clear  &  103  &  3.1  &  1.00  \\
          & 2018, Feb 21  &  STELLA               &  r'     &  189  &  1.3  &  1.86  \\
          & 2018, Apr 3   &  STELLA               &  r'     &  109  &  1.4  &  1.00  \\
\hline
\noalign{\smallskip}
HAT-P-38b & 2013, Oct 26  &  ETD, P. Benni        &  Clear  &  325  &   3.4  &  1.10  \\
          & 2013, Dec 6   &  ETD, F. Garcia       &  Clear  &  227  &   2.8  &  1.00  \\
          & 2016, Oct 29  &  ETD, M. Bretton      &  Clear  &  137  &   1.3  &  1.52  \\
          & 2017, Sep 29  &  CalarAlto2.2m        &  R      &  286  &   2.4  &  1.28  \\
\hline
\noalign{\smallskip}
HAT-P-42b & 2016, Dec 31  &  ETD, F. Lomoz        &  Clear  &  322  &   6.5  &  1.33  \\
          & 2016, Dec 31  &  ETD, F. Lomoz        &  Clear  &  367  &   7.5  &  1.04  \\
\hline
\noalign{\smallskip}
HAT-P-43b & 2014, Mar 7   &  ETD, P. Evans        &  Clear  &  200  &  5.7   &  1.11  \\
          & 2014, Nov 22  &  ETD, P. Benni        &  Clear  &  432  &  7.3   &  1.05  \\
          & 2015, Jan 17  &  ETD, J. Lozano       &  Clear  &  119  &  5.4   &  1.81  \\
          & 2016, Mar 2   &  ETD, D. Molina       &  Clear  &   92  &  6.7   &  1.32  \\
          & 2017, Dec 26  &  ETD, F. Lomoz        &  Clear  &  348  &  7.1   &  1.41  \\
          & 2017, Dec 26  &  ETD, F. Lomoz        &  Clear  &  315  & 10.0   &  1.07  \\
\hline
\noalign{\smallskip}
HAT-P-44b & 2015, Apr 21  &  ETD, M. Salisbury    &  R      &  216  &  3.4  &  1.07  \\
          & 2015, Jun 15  &  ETD, M. Salisbury    &  R      &  112  &  3.5  &  1.32  \\
          & 2016, Mar 12  &  ETD, M. Bretton      &  Clear  &   93  &  3.8  &  1.70  \\
          & 2016, Jun 6   &  ETD, A. Marchini     &  R      &   58  &  3.8  &  1.00  \\
          & 2016, Jun 6   &  ETD, M. Raetz        &  Clear  &  164  &  2.5  &  1.31  \\
          & 2017, Feb 15  &  NYSC 1m              &  R      &   53  &  1.1  &  1.28  \\ 
          & 2018, Apr 21  &  ETD, Y. Jongen       &  Clear  &   66  &  2.5  &  1.17  \\ 
          & 2018, Jun 3   &  ETD, Y. Ogmen        &  Clear  &  218  &  3.0  &  1.11  \\ 
\hline
\noalign{\smallskip}
HAT-P-45b & 2016, Jul 15  &  ETD, D. Molina       &  Clear  &   99  &  3.7  &  1.15  \\  
          & 2016, Jul 15  &  ETD, E. Diez Alonso  &  Clear  &   27  &  3.7  &  1.09  \\  
          & 2018, May 28  &  STELLA               &  r'     &   93  &  1.9  &  1.11  \\    
          & 2018, Jun 16  &  STELLA               &  r'     &   64  &  2.1  &  1.00  \\  
          & 2018, Jul 14  &  TJO                  &  V      &  347  &  4.6  &  1.03  \\
          & 2018, Aug 05  &  STELLA               &  r'     &  155  &  2.1  &  1.52  \\
\hline
\noalign{\smallskip}
HAT-P-46b & 2017, Jun 14  &  ETD, F. Lomoz        &  Clear  &  341  &  5.1  &  1.31  \\
          & 2018, Jun 15  &  TJO                  &  V      &  286  &  7.6  &  1.06  \\
          & 2018, Jun 15  &  ETD, M. Bretton      &  I      &   90  &  1.7  &  1.36  \\
          & 2018, Jun 15  &  ETD, Y. Jongen       &  Clear  &   98  &  2.6  &  1.23  \\
\hline
\noalign{\smallskip}
HAT-P-52b & 2015, Oct 16  &  ETD, P. Farissier    &  R      &   57  &  2.7  & 1.00   \\
          & 2015, Nov 9   &  ETD, M. Bretton      &  Clear  &   43  &  3.9  & 1.18   \\
          & 2016, Jan 12  &  ETD, P. Benni        &  Clear  &  389  &  8.3  & 1.00   \\
          & 2016, Jan 25  &  ETD, M. Bretton      &  Clear  &   71  &  1.9  & 1.00   \\
          & 2016, Dec 4   &  ETD, M. Bretton      &  Clear  &  102  &  1.9  & 1.66   \\
          & 2016, Dec 26  &  ETD, M. Bretton      &  Clear  &   91  &  4.5  & 1.39   \\
          & 2017, Nov 2   &  ETD, M. Bretton      &  I      &   97  &  4.0  & 1.35   \\  
          & 2018, Jan 21  &  ETD, D. Molina       &  Clear  &   57  &  4.6  & 1.00   \\  
          & 2018, Jan 21  &  ETD, F. Campos       &  Clear  &   40  &  2.3  & 1.00   \\  
\hline
\noalign{\smallskip}
KELT-3b   & 2013, Jan 4   &  ETD, R. Naves        &  R      &  215  &  2.9  & 1.00   \\
          & 2013, Mar 9   &  ETD, A. Ayiomamitis  &  Clear  &  274  &  2.6  & 1.08   \\
          & 2013, Dec 13  &  ETD, P. Benni        &  Clear  &  879  &  4.1  & 1.33   \\
          & 2015, Feb 27  &  ETD, M. Salisbury    &  R      &  579  &  3.8  & 1.25   \\
          & 2015, Apr 13  &  ETD, M. Bretton      &  V      &  358  &  2.3  & 1.24   \\
          & 2015, Dec 2   &  ETD, S. Shadick      &  I      &  259  &  4.4  & 1.14   \\
          & 2016, Mar 24  &  ETD, D. Molina       &  Clear  &  228  &  2.7  & 1.25   \\
\hline
\noalign{\smallskip}
KELT-8b   & 2016, Jul 4   &  ETD, F. Lomoz        &  B      &  380  &  7.7  & 1.35   \\  
          & 2017, Oct 11  &  STELLA               &  r'     &  174  &  4.0  & 1.04   \\  
          & 2018, May 13  &  STELLA               &  r'     &  125  &  2.4  & 1.97   \\  
          & 2018, May 26  &  STELLA               &  r'     &  172  &  1.8  & 1.12   \\  
          & 2018, Jun 8   &  STELLA               &  r'     &  113  &  3.2  & 1.40   \\  
          & 2018, Jun 11  &  STELLA               &  r'     &  186  &  1.9  & 1.90   \\ 
          & 2018, Jun 24  &  STELLA               &  r'     &  272  &  2.2  & 2.18   \\
\hline
\noalign{\smallskip}
Qatar-3b  & 2016, Dec 23  &  ETD, W. Czech        &  Clear  & 147   &  3.7  &  1.36  \\
          & 2016, Dec 23  &  ETD, M. Bretton      &  Clear  & 120   &  1.9  &  1.92  \\
          & 2016, Dec 23  &  ETD, M. Bretton      &  Clear  &  98   &  2.6  &  1.27  \\
          & 2016, Dec 28  &  ETD, Suricate48      &  Clear  &  78   &  3.4  &  1.00  \\
          & 2017, Jul 30  &  ETD, M. Morales      &  Clear  & 143   &  3.2  &  1.13  \\  
          & 2017, Nov 19  &  ETD, P. Guerra       &  Clear  & 179   &  2.9  &  1.01  \\   
\hline
\noalign{\smallskip}
Qatar-4b  & 2016, Dec 21  &  ETD, M. Bretton      &  Clear  &  156  &  2.9  &  2.41  \\
          & 2016, Dec 30  &  ETD, M. Bachschmidt  &  Clear  &  100  &  3.1  &  1.00  \\
          & 2016, Dec 30  &  ETD, M. Bretton      &  Clear  &  104  &  2.5  &  2.24  \\
          & 2017, Jan 8   &  ETD, F. Garcia       &  Clear  &   48  &  3.1  &  1.00  \\  
          & 2017, Oct 22  &  ETD, M. Salisbury    &  R      &   43  &  1.8  &  1.00  \\  
          & 2017, Oct 31  &  ETD, V.-P. Hentunen  &  R      &  146  &  6.4  &  1.70  \\  
          & 2017, Oct 31  &  ETD, V.-P. Hentunen  &  Clear  &  136  &  9.2  &  1.30  \\  
          & 2018, Jan 24  &  ETD, M. Bretton      &  I      &   72  &  2.0  &  1.35  \\  
          & 2018, Jan 24  &  ETD, F. Campos       &  Clear  &   42  &  3.1  &  1.00  \\
          & 2018, Jul 16  &  TJO                  &  V      &  307  &  4.7  &  1.25  \\

\hline
\noalign{\smallskip}
Qatar-5b  & 2016, Dec 28  &  ETD, M. Bachschmidt  &  Clear  &  131  &  5.9  &  1.23  \\
          & 2016, Dec 28  &  ETD, M. Bretton      &  Clear  &  104  &  3.8  &  1.37  \\
          & 2017, Aug 21  &  ETD, M. Bretton      &  I      &  136  &  1.8  &  1.24  \\
          & 2017, Oct 4   &  ETD, K. Fenzl        &  R      &  243  &  3.5  &  1.31  \\
          & 2018, Aug 10  &  STELLA               &  r'     &   46  &  2.4  &  1.00  \\
          & 2018, Aug 10  &  TJO                  &  V      &  126  &  2.5  &  1.11  \\
\hline
\noalign{\smallskip}
WASP-37b  & 2011, Apr 9   &  ETD, J.A. Carrion    &  R      &  124  &  5.1  & 1.44   \\
          & 2011, Apr 23  &  ETD, K. Hose         &  R      &  186  &  5.3  & 1.00   \\
          & 2011, May 11  &  ETD, S. Shadick      &  Clear  &  201  &  6.4  & 1.52   \\
          & 2012, May 3   &  ETD, A. Carreno      &  Clear  &  169  &  3.6  & 1.20   \\
          & 2013, Jun 11  &  ETD, R. Majewski     &  Clear  &   57  &  6.1  & 1.00   \\
          & 2015, Apr 22  &  ETD, J. Trnka        &  Clear  &  274  &  8.6  & 1.02   \\
          & 2018, Apr 17  &  STELLA               &  r'     &  137  &  1.5  & 1.18   \\
          & 2018, May 5   &  STELLA               &  r'     &  154  &  1.7  & 2.03   \\
\hline
\noalign{\smallskip}
WASP-58b  & 2013, Jul 14  &  ETD, J. Mravik       &  Clear  &  125  &  5.1  & 1.00   \\
          & 2013, Jul 24  &  ETD, A. Ayiomamitis  &  Clear  &  273  &  4.2  & 1.00   \\
          & 2013, Aug 19  &  ETD, F.G. Horta      &  V      &  123  &  5.4  & 1.58   \\
          & 2013, Aug 24  &  ETD, J.L. Martin     &  V      &  251  &  4.8  & 1.04   \\
          & 2015, Apr 8   &  ETD, M. Bretton      &  R      &  338  &  8.4  & 1.21   \\
          & 2016, Sep 5   &  ETD, V.-P. Hentunen  &  R      &  243  &  2.3  & 1.11   \\
          & 2017, Aug 2   &  ETD, M. Bretton      &  I      &  232  &  1.6  & 1.53   \\ 
          & 2017, Aug 2   &  ETD, R. Ballet       &  Clear  &  304  &  3.1  & 1.10   \\ 
          & 2018, May 20  &  ETD, M. Hoecherl     &  V      &  295  &  5.4  & 1.38   \\
\hline
\noalign{\smallskip}
WASP-73b  & 2018, Aug 01  &  JST                  &  R      &  1177 &  3.1  & 1.75   \\
\hline
\noalign{\smallskip}
WASP-117b & 2017, Jul 12  &  CHAT                 &  i'     &  215  &  2.1  & 1.24   \\  
          & 2017, Jul 22  &  CHAT                 &  i'     &  184  &  1.7  & 1.00   \\  
          & 2017, Jul 22  &  LCO                  &  i'     &  109  &  1.0  & 1.00   \\  
\hline                                                                                                     
\end{longtable}
}% End longtab

\longtab[2]{
\begin{longtable}{lcrc}
\caption{\label{tab_timing}Observed transit times of the investigated planets.}\\
\hline
\hline
\noalign{\smallskip}
Planet  &  BJD(TDB)  &  Epoch &  Reference \\
        &(2,450,000+) &       &       \\
\hline
\noalign{\smallskip}
\endfirsthead
\caption{Continued.} \\
\hline
\noalign{\smallskip}
Planet  &  BJD(TDB)  &  Epoch &  Reference \\
\hline
\noalign{\smallskip}
\endhead
\hline
\endfoot
\hline
\endlastfoot
HAT-P-25b &  5176.85173 $\pm$ 0.00047 &  -501  &   \cite{Quinn2012} \\
          &  6561.26872 $\pm$ 0.00066 &  -122  &   \cite{Wang2018}  \\
          &  6583.18773 $\pm$ 0.00105 &  -116  &   \cite{Wang2018}  \\
          &  6601.45096 $\pm$ 0.00097 &  -111  &   this work        \\
          &  6616.06236 $\pm$ 0.00106 &  -107  &   \cite{Wang2018}  \\
          &  6627.02128 $\pm$ 0.00058 &  -104  &   \cite{Wang2018}  \\
          &  7058.05061 $\pm$ 0.00119 &    14  &   \cite{Wang2018}  \\
          &  7405.06961 $\pm$ 0.00125 &   109  &   \cite{Wang2018}  \\
          &  7412.37504 $\pm$ 0.00105 &   111  &   this work        \\
          &  7416.02949 $\pm$ 0.00178 &   112  &   \cite{Wang2018}  \\
          &  7664.41978 $\pm$ 0.00068 &   180  &   this work        \\
          &  7697.29563 $\pm$ 0.00065 &   189  &   \cite{Wang2018}  \\
          &  8095.45305 $\pm$ 0.00064 &   298  &   this work        \\
          &  8106.40933 $\pm$ 0.00062 &   301  &   this work        \\
\hline
\noalign{\smallskip}
HAT-P-29b &   5563.87156 $\pm$ 0.00065  & -267  & \cite{Wang2018b} \\
          &   5586.76257 $\pm$ 0.00061  & -263  & \cite{Wang2018b} \\
          &   5838.59183 $\pm$ 0.00325  & -219  &   this work  \\
          &   6193.43030 $\pm$ 0.00402  & -157  &   this work  \\
          &   6691.36248 $\pm$ 0.00266  &  -70  &   this work  \\
          &   8076.42237 $\pm$ 0.00244  &  172  &   this work  \\
          &   8116.48552 $\pm$ 0.00126  &  179  &   this work  \\
\hline
\noalign{\smallskip}
HAT-P-31b &   4320.8866 $\pm$ 0.0051   & -769  &  \cite{Kipping2011}  \\
          &   8270.05094 $\pm$ 0.00564 &   20  &   this work  \\    
          &   8320.09907 $\pm$ 0.00131 &   30  &   this work  \\
          &   8320.09673 $\pm$ 0.00550 &   30  &   this work  \\
          &   8330.10726 $\pm$ 0.00340 &   32  &   this work  \\
          &   8335.11829 $\pm$ 0.00213 &   33  &   this work  \\
\hline
\noalign{\smallskip}
HAT-P-34b & 5431.59706 $\pm$ 0.00055  &  -189  &  \cite{Bakos2012} \\
          & 6156.78867 $\pm$ 0.00313  &   -56  &  this work \\
          & 6478.49205 $\pm$ 0.00136  &     3  &  this work \\
          & 7923.44147 $\pm$ 0.00381  &   268  &  this work \\
          & 7983.42395 $\pm$ 0.00406  &   279  &  this work \\
          & 8261.50682 $\pm$ 0.00183  &   330  &  this work \\
          & 8310.60566 $\pm$ 0.00140  &   339  &  this work \\
\hline
\noalign{\smallskip}
HAT-P-35b & 5578.66081  $\pm$ 0.00050 &  -345  &   \cite{Bakos2012} \\
          & 7449.39606  $\pm$ 0.00207 &   168  &   this work    \\
          & 8171.43644  $\pm$ 0.00127 &   366  &   this work    \\
          & 8211.54620  $\pm$ 0.00118 &   377  &   this work    \\
\hline
\noalign{\smallskip}
HAT-P-38b &  5863.11957 $\pm$ 0.00035 &  -351  &   \cite{Sato2012} \\
          &  6591.65204 $\pm$ 0.00139 &  -194  &   this work  \\
          &  6633.41297 $\pm$ 0.00106 &  -185  &   this work  \\
          &  7450.11375 $\pm$ 0.00045 &    -9  &   \cite{Bruno2018} \\
          &  7626.44542 $\pm$ 0.00010 &    29  &   \cite{Bruno2018} \\
          &  7691.40793 $\pm$ 0.00103 &    43  &   this work  \\
          &  8025.51268 $\pm$ 0.00149 &   115  &   this work  \\
\hline
\noalign{\smallskip}
HAT-P-42b &  5952.52683 $\pm$ 0.00077 &  -18  &  \cite{Boisse2013} \\
          &  7753.55133 $\pm$ 0.00639 &  370  &   this work \\
          &  7753.56116 $\pm$ 0.00417 &  370  &   this work \\
\hline
\noalign{\smallskip}
HAT-P-43b &  5997.37182 $\pm$ 0.00032 &  -45  &   \cite{Boisse2013} \\
          &  6723.89545 $\pm$ 0.00162 &  173  &   this work \\
          &  6983.84571 $\pm$ 0.00148 &  251  &   this work \\
          &  7040.49786 $\pm$ 0.00287 &  268  &   this work \\
          &  7450.41836 $\pm$ 0.00345 &  391  &   this work \\
          &  8113.62728 $\pm$ 0.00255 &  590  &   this work \\
          &  8113.62975 $\pm$ 0.00236 &  590  &   this work \\
\hline
\noalign{\smallskip}
HAT-P-44b & 5696.93772 $\pm$ 0.00024  &  -118  &   \cite{Hartman2014} \\
          & 7133.53528 $\pm$ 0.00100  &   216  &   this work  \\
          & 7189.44845 $\pm$ 0.00226  &   229  &   this work  \\
          & 7460.42717 $\pm$ 0.00194  &   292  &   this work  \\
          & 7546.45090 $\pm$ 0.00172  &   312  &   this work  \\
          & 7546.44706 $\pm$ 0.00127  &   312  &   this work  \\
          & 7800.21697 $\pm$ 0.00083  &   371  &   this work  \\
          & 8230.34088 $\pm$ 0.00180  &   471  &   this work  \\
          & 8273.35095 $\pm$ 0.00081  &   481  &   this work  \\
\hline
\noalign{\smallskip}
HAT-P-45b & 5729.98689 $\pm$ 0.00041  &  -247  &   \cite{Hartman2014} \\
          & 7585.47810 $\pm$ 0.00222  &   346  &   this work  \\
          & 7585.47804 $\pm$ 0.00319  &   346  &   this work  \\
          & 8267.59965 $\pm$ 0.00148  &   564  &   this work  \\
          & 8286.37230 $\pm$ 0.00120  &   570  &   this work  \\
          & 8314.53462 $\pm$ 0.00110  &   579  &   this work  \\
          & 8336.43982 $\pm$ 0.00159  &   586  &   this work  \\
\hline
\noalign{\smallskip}
HAT-P-46b & 5701.33723 $\pm$ 0.00047  &  -60  &   \cite{Hartman2014} \\
          & 7919.51715 $\pm$ 0.00326  &  437  &   this work  \\
          & 8285.49459 $\pm$ 0.00451  &  519  &   this work  \\
          & 8285.49325 $\pm$ 0.00184  &  519  &   this work  \\
          & 8285.49321 $\pm$ 0.00312  &  519  &   this work  \\
\hline
\noalign{\smallskip}
HAT-P-52b & 5852.10403 $\pm$ 0.00041  &  -288  &  \cite{Hartman2015_52} \\
          & 7311.51042 $\pm$ 0.00142  &   242  &  this work \\
          & 7336.29302 $\pm$ 0.00277  &   251  &  this work \\
          & 7399.62421 $\pm$ 0.00133  &   274  &  this work \\
          & 7413.39279 $\pm$ 0.00091  &   279  &  this work \\
          & 7727.30439 $\pm$ 0.00107  &   393  &  this work \\
          & 7749.32972 $\pm$ 0.00193  &   401  &  this work \\
          & 8060.48649 $\pm$ 0.00155  &   514  &  this work \\
          & 8140.34636 $\pm$ 0.00214  &   543  &  this work \\
          & 8140.34268 $\pm$ 0.00137  &   543  &  this work \\
\hline
\noalign{\smallskip}
KELT-3b   & 6034.29537 $\pm$ 0.00038  &  -87  &  \cite{Pepper2013} \\
          & 6296.52093 $\pm$ 0.00162  &   10  &  this work \\
          & 6361.40497 $\pm$ 0.00127  &   34  &  this work \\
          & 6639.85526 $\pm$ 0.00230  &  137  &  this work \\
          & 7080.50825 $\pm$ 0.00205  &  300  &  this work \\
          & 7126.46210 $\pm$ 0.00106  &  317  &  this work \\
          & 7358.95516 $\pm$ 0.00240  &  403  &  this work \\
          & 7472.49400 $\pm$ 0.00188  &  445  &  this work \\
\hline
\noalign{\smallskip}
KELT-8b   & 6883.4803  $\pm$ 0.0007    &  -158  &  \cite{Fulton2015} \\
          & 7574.47027 $\pm$ 0.00464   &    55  & this work \\ 
          & 8038.37293 $\pm$ 0.00268   &   198  & this work \\ 
          & 8252.4856 $\pm$  0.0076    &   264  & this work \\ 
          & 8265.45835 $\pm$ 0.00121   &   268  & this work \\ 
          & 8278.43966 $\pm$ 0.00446   &   272  & this work \\ 
          & 8281.67389 $\pm$ 0.00234   &   273  & this work \\
          & 8294.65635 $\pm$ 0.00258   &   277  & this work \\
\hline
\noalign{\smallskip}
Qatar-3b  & 7302.45300 $\pm$ 0.00010  &   -4  &   \cite{Alsubai2017} \\
          & 7746.34721 $\pm$ 0.00260  &  173  &   this work  \\
          & 7746.35109 $\pm$ 0.00272  &  173  &   this work  \\
          & 7746.35313 $\pm$ 0.00201  &  173  &   this work  \\
          & 7751.36766 $\pm$ 0.00214  &  175  &   this work  \\
          & 7964.53543 $\pm$ 0.00165  &  260  &   this work  \\
          & 8077.39371 $\pm$ 0.00132  &  305  &   this work  \\
\hline
\noalign{\smallskip}
Qatar-4b  & 7637.77361 $\pm$ 0.00046  &  -130  &   \cite{Alsubai2017} \\
          & 7744.28950 $\pm$ 0.00164  &   -71  &   this work  \\
          & 7753.31672 $\pm$ 0.00074  &   -66  &   this work  \\
          & 7753.31884 $\pm$ 0.00115  &   -66  &   this work  \\
          & 7762.34209 $\pm$ 0.00125  &   -61  &   this work  \\
          & 8049.39801 $\pm$ 0.00058  &   103  &   this work  \\
          & 8058.42511 $\pm$ 0.00203  &   103  &   this work  \\
          & 8058.42459 $\pm$ 0.00235  &    98  &   this work  \\
          & 8143.27752 $\pm$ 0.00082  &   150  &   this work  \\
          & 8143.27761 $\pm$ 0.00159  &   150  &   this work  \\
          & 8316.59101 $\pm$ 0.00103  &   246  &   this work  \\
\hline
\noalign{\smallskip}
Qatar-5b  & 7336.75824 $\pm$ 0.00010  &   -9  &   \cite{Alsubai2017} \\
          & 7751.37715 $\pm$ 0.00338  &  135  &   this work  \\
          & 7751.38365 $\pm$ 0.00240  &  135  &   this work  \\
          & 7987.48158 $\pm$ 0.00074  &  217  &   this work  \\
          & 8030.67394 $\pm$ 0.00113  &  232  &   this work  \\
          & 8341.63656 $\pm$ 0.00154  &  340  &   this work  \\
          & 8341.63779 $\pm$ 0.00127  &  340  &   this work  \\
\hline
\noalign{\smallskip}
WASP-37b  & 5338.6196 $\pm$ 0.0006    &  -295  &   \cite{Simpson2011} \\
          & 5660.59180 $\pm$ 0.00284  &  -205  &  this work  \\
          & 5674.90434 $\pm$ 0.00138  &  -201  &  this work  \\
          & 5692.79172 $\pm$ 0.00267  &  -196  &  this work  \\
          & 6050.53938 $\pm$ 0.00145  &   -96  &  this work  \\
          & 6454.79476 $\pm$ 0.00348  &    17  &  this work  \\
          & 7134.51870 $\pm$ 0.00211  &   207  &  this work  \\
          & 8225.64753 $\pm$ 0.00892  &   512  &  this work  \\
          & 8243.53184 $\pm$ 0.00125  &   517  &  this work  \\
\hline
\noalign{\smallskip}
WASP-58b  & 5183.9342 $\pm$ 0.0010    &  -414  &  \cite{Hebrard2013} \\
          & 6488.40794 $\pm$ 0.00264  &  -154  &  this work \\
          & 6498.44187 $\pm$ 0.00121  &  -152  &  this work \\
          & 6523.52545 $\pm$ 0.00316  &  -147  &  this work \\
          & 6528.54704 $\pm$ 0.00134  &  -146  &  this work \\
          & 7120.57537 $\pm$ 0.00297  &   -28  &  this work \\
          & 7637.35161 $\pm$ 0.00089  &    75  &  this work \\
          & 7968.48759 $\pm$ 0.00068  &   141  &  this work \\
          & 7968.48541 $\pm$ 0.00082  &   141  &  this work \\
          & 8259.48221 $\pm$ 0.00249  &   199  &  this work \\
\hline
\noalign{\smallskip}
WASP-73b  & 6128.7063 $\pm$ 0.0011  &  -58   &  \cite{Delrez2014} \\
          & 8331.7531 $\pm$ 0.0045  &  481  &  this work \\
\hline
\noalign{\smallskip}
WASP-117b & 6533.82404 $\pm$ 0.00095  &  -82  &  \cite{Lendl2014} \\
          & 7946.72810 $\pm$ 0.00198  &   59  &  this work \\
          & 7956.74985 $\pm$ 0.00163  &   60  &  this work \\
          & 7956.75113 $\pm$ 0.00105  &   60  &  this work \\

\hline                                                                                                     
\end{longtable}
}% End longtab

\end{appendix}

\end{document}